\documentclass[11pt,twoside,a4paper,cmspaper,final,collab]{cms-tdr}

\pdfoutput=1
\def\svnVersion{13:8673M}\def\cmsCernNo{2010-014}\def\cmsCernDate{\today}

\begin{document}\cmsNoteHeader{QCD-10-001}
%
%
%

%
%
\hyphenation{env-iron-men-tal}
\hyphenation{had-ron-i-za-tion}
\hyphenation{cal-or-i-me-ter}
\hyphenation{de-vices}
%
\RCS$Revision: 7825 $
\RCS$HeadURL: svn+ssh://alverson@svn.cern.ch/reps/tdr2/papers/QCD-10-001/trunk/QCD-10-001.tex $
\RCS$Id: QCD-10-001.tex 7825 2010-06-05 19:27:18Z pmarage $


%
%
%

\providecommand {\etal}{\mbox{et al.}\xspace} 
\providecommand {\ie}{\mbox{i.e.}\xspace}     
\providecommand {\eg}{\mbox{e.g.}\xspace}     
\providecommand {\etc}{\mbox{etc.}\xspace}     
\providecommand {\vs}{\mbox{\sl vs.}\xspace}      
\providecommand {\mdash}{\ensuremath{\mathrm{-}}} 

\providecommand {\Lone}{Level-1\xspace} 
\providecommand {\Ltwo}{Level-2\xspace}
\providecommand {\Lthree}{Level-3\xspace}

\providecommand{\ACERMC} {\textsc{AcerMC}\xspace}
\providecommand{\ALPGEN} {{\textsc{alpgen}}\xspace}
\providecommand{\CHARYBDIS} {{\textsc{charybdis}}\xspace}
\providecommand{\CMKIN} {\textsc{cmkin}\xspace}
\providecommand{\CMSIM} {{\textsc{cmsim}}\xspace}
\providecommand{\CMSSW} {{\textsc{cmssw}}\xspace}
\providecommand{\COBRA} {{\textsc{cobra}}\xspace}
\providecommand{\COCOA} {{\textsc{cocoa}}\xspace}
\providecommand{\COMPHEP} {\textsc{CompHEP}\xspace}
\providecommand{\EVTGEN} {{\textsc{evtgen}}\xspace}
\providecommand{\FAMOS} {{\textsc{famos}}\xspace}
\providecommand{\GARCON} {\textsc{garcon}\xspace}
\providecommand{\GARFIELD} {{\textsc{garfield}}\xspace}
\providecommand{\GEANE} {{\textsc{geane}}\xspace}
\providecommand{\GEANTfour} {{\textsc{geant4}}\xspace}
\providecommand{\GEANTthree} {{\textsc{geant3}}\xspace}
\providecommand{\GEANT} {{\textsc{geant}}\xspace}
\providecommand{\HDECAY} {\textsc{hdecay}\xspace}
\providecommand{\HERWIG} {{\textsc{herwig}}\xspace}
\providecommand{\HIGLU} {{\textsc{higlu}}\xspace}
\providecommand{\HIJING} {{\textsc{hijing}}\xspace}
\providecommand{\IGUANA} {\textsc{iguana}\xspace}
\providecommand{\ISAJET} {{\textsc{isajet}}\xspace}
\providecommand{\ISAPYTHIA} {{\textsc{isapythia}}\xspace}
\providecommand{\ISASUGRA} {{\textsc{isasugra}}\xspace}
\providecommand{\ISASUSY} {{\textsc{isasusy}}\xspace}
\providecommand{\ISAWIG} {{\textsc{isawig}}\xspace}
\providecommand{\MADGRAPH} {\textsc{MadGraph}\xspace}
\providecommand{\MCATNLO} {\textsc{mc@nlo}\xspace}
\providecommand{\MCFM} {\textsc{mcfm}\xspace}
\providecommand{\MILLEPEDE} {{\textsc{millepede}}\xspace}
\providecommand{\ORCA} {{\textsc{orca}}\xspace}
\providecommand{\OSCAR} {{\textsc{oscar}}\xspace}
\providecommand{\PHOTOS} {\textsc{photos}\xspace}
\providecommand{\PROSPINO} {\textsc{prospino}\xspace}
\providecommand{\PYTHIA} {{\textsc{pythia}}\xspace}
\providecommand{\SHERPA} {{\textsc{sherpa}}\xspace}
\providecommand{\TAUOLA} {\textsc{tauola}\xspace}
\providecommand{\TOPREX} {\textsc{TopReX}\xspace}
\providecommand{\XDAQ} {{\textsc{xdaq}}\xspace}

\providecommand {\DZERO}{D\O\xspace}     


\providecommand{\de}{\ensuremath{^\circ}}
\providecommand{\ten}[1]{\ensuremath{\times \text{10}^\text{#1}}}
\providecommand{\unit}[1]{\ensuremath{\text{\,#1}}\xspace}
\providecommand{\mum}{\ensuremath{\,\mu\text{m}}\xspace}
\providecommand{\micron}{\ensuremath{\,\mu\text{m}}\xspace}
\providecommand{\cm}{\ensuremath{\,\text{cm}}\xspace}
\providecommand{\mm}{\ensuremath{\,\text{mm}}\xspace}
\providecommand{\mus}{\ensuremath{\,\mu\text{s}}\xspace}
\providecommand{\keV}{\ensuremath{\,\text{ke\hspace{-.08em}V}}\xspace}
\providecommand{\MeV}{\ensuremath{\,\text{Me\hspace{-.08em}V}}\xspace}
\providecommand{\GeV}{\ensuremath{\,\text{Ge\hspace{-.08em}V}}\xspace}
\providecommand{\TeV}{\ensuremath{\,\text{Te\hspace{-.08em}V}}\xspace}
\providecommand{\PeV}{\ensuremath{\,\text{Pe\hspace{-.08em}V}}\xspace}
\providecommand{\keVc}{\ensuremath{{\,\text{ke\hspace{-.08em}V\hspace{-0.16em}/\hspace{-0.08em}}c}}\xspace}
\providecommand{\MeVc}{\ensuremath{{\,\text{Me\hspace{-.08em}V\hspace{-0.16em}/\hspace{-0.08em}}c}}\xspace}
\providecommand{\GeVc}{\ensuremath{{\,\text{Ge\hspace{-.08em}V\hspace{-0.16em}/\hspace{-0.08em}}c}}\xspace}
\providecommand{\TeVc}{\ensuremath{{\,\text{Te\hspace{-.08em}V\hspace{-0.16em}/\hspace{-0.08em}}c}}\xspace}
\providecommand{\keVcc}{\ensuremath{{\,\text{ke\hspace{-.08em}V\hspace{-0.16em}/\hspace{-0.08em}}c^\text{2}}}\xspace}
\providecommand{\MeVcc}{\ensuremath{{\,\text{Me\hspace{-.08em}V\hspace{-0.16em}/\hspace{-0.08em}}c^\text{2}}}\xspace}
\providecommand{\GeVcc}{\ensuremath{{\,\text{Ge\hspace{-.08em}V\hspace{-0.16em}/\hspace{-0.08em}}c^\text{2}}}\xspace}
\providecommand{\TeVcc}{\ensuremath{{\,\text{Te\hspace{-.08em}V\hspace{-0.16em}/\hspace{-0.08em}}c^\text{2}}}\xspace}

\providecommand{\pbinv} {\mbox{\ensuremath{\,\text{pb}^\text{$-$1}}}\xspace}
\providecommand{\fbinv} {\mbox{\ensuremath{\,\text{fb}^\text{$-$1}}}\xspace}
\providecommand{\nbinv} {\mbox{\ensuremath{\,\text{nb}^\text{$-$1}}}\xspace}
\providecommand{\percms}{\ensuremath{\,\text{cm}^\text{$-$2}\,\text{s}^\text{$-$1}}\xspace}
\providecommand{\lumi}{\ensuremath{\mathcal{L}}\xspace}
\providecommand{\Lumi}{\ensuremath{\mathcal{L}}\xspace}
%
\providecommand{\LvLow}  {\ensuremath{\mathcal{L}=\text{10}^\text{32}\,\text{cm}^\text{$-$2}\,\text{s}^\text{$-$1}}\xspace}
\providecommand{\LLow}   {\ensuremath{\mathcal{L}=\text{10}^\text{33}\,\text{cm}^\text{$-$2}\,\text{s}^\text{$-$1}}\xspace}
\providecommand{\lowlumi}{\ensuremath{\mathcal{L}=\text{2}\times \text{10}^\text{33}\,\text{cm}^\text{$-$2}\,\text{s}^\text{$-$1}}\xspace}
\providecommand{\LMed}   {\ensuremath{\mathcal{L}=\text{2}\times \text{10}^\text{33}\,\text{cm}^\text{$-$2}\,\text{s}^\text{$-$1}}\xspace}
\providecommand{\LHigh}  {\ensuremath{\mathcal{L}=\text{10}^\text{34}\,\text{cm}^\text{-2}\,\text{s}^\text{$-$1}}\xspace}
\providecommand{\hilumi} {\ensuremath{\mathcal{L}=\text{10}^\text{34}\,\text{cm}^\text{-2}\,\text{s}^\text{$-$1}}\xspace}


\providecommand{\zp}{\ensuremath{\mathrm{Z}^\prime}\xspace}


\providecommand{\kt}{\ensuremath{k_{\mathrm{T}}}\xspace}
\providecommand{\BC}{\ensuremath{{B_{\mathrm{c}}}}\xspace}
\providecommand{\bbarc}{\ensuremath{{\overline{\mathrm{b}}\mathrm{c}}}\xspace}
\providecommand{\bbbar}{\ensuremath{{\mathrm{b}\overline{\mathrm{b}}}}\xspace}
\providecommand{\ccbar}{\ensuremath{{\mathrm{c}\overline{\mathrm{c}}}}\xspace}
\providecommand{\JPsi}{\ensuremath{{\mathrm{J}}\hspace{-.08em}/\hspace{-.14em}\psi}\xspace}
\providecommand{\bspsiphi}{\ensuremath{\mathrm{B}_\mathrm{s} \to \JPsi\, \phi}\xspace}
\providecommand{\AFB}{\ensuremath{A_\text{FB}}\xspace}
\providecommand{\EE}{\ensuremath{\mathrm{e}^+\mathrm{e}^-}\xspace}
\providecommand{\MM}{\ensuremath{\mu^+\mu^-}\xspace}
\providecommand{\TT}{\ensuremath{\tau^+\tau^-}\xspace}
\providecommand{\wangle}{\ensuremath{\sin^{2}\theta_{\text{eff}}^\text{lept}(M^2_\mathrm{Z})}\xspace}
\providecommand{\ttbar}{\ensuremath{{\mathrm{t}\overline{\mathrm{t}}}}\xspace}
\providecommand{\stat}{\ensuremath{\,\text{(stat.)}}\xspace}
\providecommand{\syst}{\ensuremath{\,\text{(syst.)}}\xspace}

\providecommand{\HGG}{\ensuremath{\mathrm{H}\to\gamma\gamma}}
\providecommand{\gev}{\GeV}
\providecommand{\GAMJET}{\ensuremath{\gamma + \text{jet}}}
\providecommand{\PPTOJETS}{\ensuremath{\mathrm{pp}\to\text{jets}}}
\providecommand{\PPTOGG}{\ensuremath{\mathrm{pp}\to\gamma\gamma}}
\providecommand{\PPTOGAMJET}{\ensuremath{\mathrm{pp}\to\gamma + \mathrm{jet}}}
\providecommand{\MH}{\ensuremath{\mathrm{M_{\mathrm{H}}}}}
\providecommand{\RNINE}{\ensuremath{\mathrm{R}_\mathrm{9}}}
\providecommand{\DR}{\ensuremath{\Delta\mathrm{R}}}


\providecommand{\PT}{\ensuremath{p_{\mathrm{T}}}\xspace}
\providecommand{\pt}{\ensuremath{p_{\mathrm{T}}}\xspace}
\providecommand{\ET}{\ensuremath{E_{\mathrm{T}}}\xspace}
\providecommand{\HT}{\ensuremath{H_{\mathrm{T}}}\xspace}
\providecommand{\et}{\ensuremath{E_{\mathrm{T}}}\xspace}
\providecommand{\Em}{\ensuremath{E\!\!\!/}\xspace}
\providecommand{\Pm}{\ensuremath{p\!\!\!/}\xspace}
\providecommand{\PTm}{\ensuremath{{p\!\!\!/}_{\mathrm{T}}}\xspace}
\providecommand{\ETm}{\ensuremath{E_{\mathrm{T}}^{\text{miss}}}\xspace}
\providecommand{\MET}{\ensuremath{E_{\mathrm{T}}^{\text{miss}}}\xspace}
\providecommand{\ETmiss}{\ensuremath{E_{\mathrm{T}}^{\text{miss}}}\xspace}
\providecommand{\VEtmiss}{\ensuremath{{\vec E}_{\mathrm{T}}^{\text{miss}}}\xspace}

\providecommand{\dd}[2]{\ensuremath{\frac{\mathrm{d} #1}{\mathrm{d} #2}}}

%

\providecommand{\ga}{\ensuremath{\gtrsim}}
\providecommand{\la}{\ensuremath{\lesssim}}
\providecommand{\swsq}{\ensuremath{\sin^2\theta_\mathrm{W}}\xspace}
\providecommand{\cwsq}{\ensuremath{\cos^2\theta_\mathrm{W}}\xspace}
\providecommand{\tanb}{\ensuremath{\tan\beta}\xspace}
\providecommand{\tanbsq}{\ensuremath{\tan^{2}\beta}\xspace}
\providecommand{\sidb}{\ensuremath{\sin 2\beta}\xspace}
\providecommand{\alpS}{\ensuremath{\alpha_S}\xspace}
\providecommand{\alpt}{\ensuremath{\tilde{\alpha}}\xspace}

\providecommand{\QL}{\ensuremath{Q_L}\xspace}
\providecommand{\sQ}{\ensuremath{\tilde{Q}}\xspace}
\providecommand{\sQL}{\ensuremath{\tilde{Q}_L}\xspace}
\providecommand{\ULC}{\ensuremath{U_L^C}\xspace}
\providecommand{\sUC}{\ensuremath{\tilde{U}^C}\xspace}
\providecommand{\sULC}{\ensuremath{\tilde{U}_L^C}\xspace}
\providecommand{\DLC}{\ensuremath{D_L^C}\xspace}
\providecommand{\sDC}{\ensuremath{\tilde{D}^C}\xspace}
\providecommand{\sDLC}{\ensuremath{\tilde{D}_L^C}\xspace}
\providecommand{\LL}{\ensuremath{L_L}\xspace}
\providecommand{\sL}{\ensuremath{\tilde{L}}\xspace}
\providecommand{\sLL}{\ensuremath{\tilde{L}_L}\xspace}
\providecommand{\ELC}{\ensuremath{E_L^C}\xspace}
\providecommand{\sEC}{\ensuremath{\tilde{E}^C}\xspace}
\providecommand{\sELC}{\ensuremath{\tilde{E}_L^C}\xspace}
\providecommand{\sEL}{\ensuremath{\tilde{E}_L}\xspace}
\providecommand{\sER}{\ensuremath{\tilde{E}_R}\xspace}
\providecommand{\sFer}{\ensuremath{\tilde{f}}\xspace}
\providecommand{\sQua}{\ensuremath{\tilde{q}}\xspace}
\providecommand{\sUp}{\ensuremath{\tilde{u}}\xspace}
\providecommand{\suL}{\ensuremath{\tilde{u}_L}\xspace}
\providecommand{\suR}{\ensuremath{\tilde{u}_R}\xspace}
\providecommand{\sDw}{\ensuremath{\tilde{d}}\xspace}
\providecommand{\sdL}{\ensuremath{\tilde{d}_L}\xspace}
\providecommand{\sdR}{\ensuremath{\tilde{d}_R}\xspace}
\providecommand{\sTop}{\ensuremath{\tilde{t}}\xspace}
\providecommand{\stL}{\ensuremath{\tilde{t}_L}\xspace}
\providecommand{\stR}{\ensuremath{\tilde{t}_R}\xspace}
\providecommand{\stone}{\ensuremath{\tilde{t}_1}\xspace}
\providecommand{\sttwo}{\ensuremath{\tilde{t}_2}\xspace}
\providecommand{\sBot}{\ensuremath{\tilde{b}}\xspace}
\providecommand{\sbL}{\ensuremath{\tilde{b}_L}\xspace}
\providecommand{\sbR}{\ensuremath{\tilde{b}_R}\xspace}
\providecommand{\sbone}{\ensuremath{\tilde{b}_1}\xspace}
\providecommand{\sbtwo}{\ensuremath{\tilde{b}_2}\xspace}
\providecommand{\sLep}{\ensuremath{\tilde{l}}\xspace}
\providecommand{\sLepC}{\ensuremath{\tilde{l}^\mathrm{C}}\xspace}
\providecommand{\sEl}{\ensuremath{\tilde{\mathrm{e}}}\xspace}
\providecommand{\sElC}{\ensuremath{\tilde{\mathrm{e}}^\mathrm{C}}\xspace}
\providecommand{\seL}{\ensuremath{\tilde{\mathrm{e}}_\mathrm{L}}\xspace}
\providecommand{\seR}{\ensuremath{\tilde{\mathrm{e}}_\mathrm{R}}\xspace}
\providecommand{\snL}{\ensuremath{\tilde{\nu}_L}\xspace}
\providecommand{\sMu}{\ensuremath{\tilde{\mu}}\xspace}
\providecommand{\sNu}{\ensuremath{\tilde{\nu}}\xspace}
\providecommand{\sTau}{\ensuremath{\tilde{\tau}}\xspace}
\providecommand{\Glu}{\ensuremath{g}\xspace}
\providecommand{\sGlu}{\ensuremath{\tilde{g}}\xspace}
\providecommand{\Wpm}{\ensuremath{\mathrm{W}^{\pm}}\xspace}
\providecommand{\sWpm}{\ensuremath{\tilde{\mathrm{W}}^{\pm}}\xspace}
\providecommand{\Wz}{\ensuremath{\mathrm{W}^{0}}\xspace}
\providecommand{\sWz}{\ensuremath{\tilde{\mathrm{W}}^{0}}\xspace}
\providecommand{\sWino}{\ensuremath{\tilde{\mathrm{W}}}\xspace}
\providecommand{\Bz}{\ensuremath{\mathrm{B}^{0}}\xspace}
\providecommand{\sBz}{\ensuremath{\tilde{\mathrm{B}}^{0}}\xspace}
\providecommand{\sBino}{\ensuremath{\tilde{\mathrm{B}}}\xspace}
\providecommand{\Zz}{\ensuremath{\mathrm{Z}^{0}}\xspace}
\providecommand{\sZino}{\ensuremath{\tilde{\mathrm{Z}}^{0}}\xspace}
\providecommand{\sGam}{\ensuremath{\tilde{\gamma}}\xspace}
\providecommand{\chiz}{\ensuremath{\tilde{\chi}^{0}}\xspace}
\providecommand{\chip}{\ensuremath{\tilde{\chi}^{+}}\xspace}
\providecommand{\chim}{\ensuremath{\tilde{\chi}^{-}}\xspace}
\providecommand{\chipm}{\ensuremath{\tilde{\chi}^{\pm}}\xspace}
\providecommand{\Hone}{\ensuremath{\mathrm{H}_\mathrm{d}}\xspace}
\providecommand{\sHone}{\ensuremath{\tilde{\mathrm{H}}_\mathrm{d}}\xspace}
\providecommand{\Htwo}{\ensuremath{\mathrm{H}_\mathrm{u}}\xspace}
\providecommand{\sHtwo}{\ensuremath{\tilde{\mathrm{H}}_\mathrm{u}}\xspace}
\providecommand{\sHig}{\ensuremath{\tilde{\mathrm{H}}}\xspace}
\providecommand{\sHa}{\ensuremath{\tilde{\mathrm{H}}_\mathrm{a}}\xspace}
\providecommand{\sHb}{\ensuremath{\tilde{\mathrm{H}}_\mathrm{b}}\xspace}
\providecommand{\sHpm}{\ensuremath{\tilde{\mathrm{H}}^{\pm}}\xspace}
\providecommand{\hz}{\ensuremath{\mathrm{h}^{0}}\xspace}
\providecommand{\Hz}{\ensuremath{\mathrm{H}^{0}}\xspace}
\providecommand{\Az}{\ensuremath{\mathrm{A}^{0}}\xspace}
\providecommand{\Hpm}{\ensuremath{\mathrm{H}^{\pm}}\xspace}
\providecommand{\sGra}{\ensuremath{\tilde{\mathrm{G}}}\xspace}
\providecommand{\mtil}{\ensuremath{\tilde{m}}\xspace}
\providecommand{\rpv}{\ensuremath{\rlap{\kern.2em/}R}\xspace}
\providecommand{\LLE}{\ensuremath{LL\bar{E}}\xspace}
\providecommand{\LQD}{\ensuremath{LQ\bar{D}}\xspace}
\providecommand{\UDD}{\ensuremath{\overline{UDD}}\xspace}
\providecommand{\Lam}{\ensuremath{\lambda}\xspace}
\providecommand{\Lamp}{\ensuremath{\lambda'}\xspace}
\providecommand{\Lampp}{\ensuremath{\lambda''}\xspace}
\providecommand{\spinbd}[2]{\ensuremath{\bar{#1}_{\dot{#2}}}\xspace}

\providecommand{\MD}{\ensuremath{{M_\mathrm{D}}}\xspace}
\providecommand{\Mpl}{\ensuremath{{M_\mathrm{Pl}}}\xspace}
\providecommand{\Rinv} {\ensuremath{{R}^{-1}}\xspace}

\def\lsim{\! \mathrel{\rlap{\lower4pt\hbox{\hskip1pt$\sim$}}
    \raise1pt\hbox{$<$}} \! }         
\def\gsim{\! \mathrel{\rlap{\lower4pt\hbox{\hskip1pt$\sim$}}
    \raise1pt\hbox{$>$}} \! }

\newcommand*{\py}{PYTHIA}


\newcommand*{\ptcut}{$p_T\!>\!0.5\,{\rm GeV/}c$}
\newcommand*{\etacut}{$|\eta|\!<\!2$}
\newcommand*{\delphi}{$\Delta\phi$}


\cmsNoteHeader{QCD-10-001} 

\title{First Measurement of the Underlying Event Activity \\
at the LHC  with $\sqrt{s} = 0.9~\TeV$ }       

\author[cern]{The CMS Collaboration}

\date{\today}


\abstract{
A measurement of the underlying activity in scattering processes with
$p_T$ scale in the GeV region
is performed in proton-proton collisions at $\sqrt{s} = 0.9 \TeV$, using data
collected by the CMS experiment at the LHC.
Charged hadron production is studied with reference to the direction of a
leading object, either a charged particle or a set of charged particles
forming a jet.
Predictions of several QCD-inspired models as implemented in PYTHIA are
compared, after full detector simulation, to the data. 
The models generally predict too little production of charged hadrons with
pseudorapidity \etacut, \ptcut, and azimuthal direction transverse to that of the 
leading object.
}

\hypersetup{%
pdfauthor={CMS Collaboration},%
pdftitle={Measurement of the Underlying Event Activity in Proton-Proton Collisions at 0.9 TeV},%
pdfsubject={CMS},%
pdfkeywords={CMS, QCD, physics}}

\maketitle 


\section{Introduction}
\label{sec:Introduction}

In the presence of a ``hard" process characterized by  large transverse 
momenta $p_T$ with respect to the beam direction, the hadronic final states of 
hadron-hadron interactions can be described 
as the superposition of several contributions:
products of the partonic hard scattering with the highest $p_T$, 
including initial and final state radiation;
hadrons produced in additional ``multiple parton interactions" (MPI);
and ``beam-beam remnants" (BBR) resulting from the hadronization of
the partonic constituents that did not participate in other scatters.
Products of MPI and BBR form the ``underlying event" (UE).
The UE cannot be uniquely separated from
initial and final state radiation.

A good description of UE properties is crucial for precision measurements
of Standard Model processes and the search for new physics 
at the CERN Large Hadron Collider (LHC)~\cite{Evans:2008zzb}.
Multiplicity distributions measured by the UA5 collaboration
at the ${\rm Sp \bar p S}$ collider~\cite{Alner:1986xu} were
modeled in Monte Carlo (MC) simulations~\cite{Sjostrand:1986ep}.
Detailed UE studies performed at the Tevatron by the CDF
collaboration~\cite{Affolder:2001xt,Acosta:2004wqa,NewRickCitation}
led to significant progress in MPI modeling~\cite{Bartalini:2009xx}.
The UE dynamics is, however, not fully understood, especially the
centre-of-mass energy dependence.
A new energy domain is opening with the LHC, where UE properties
can be studied with data taken at $\sqrt{s} = 0.9$, $7$, and $14 \TeV$.
The data at $0.9 \TeV$ analyzed in this paper provide a valuable reference 
point to progress in the understanding of UE and MPI.

UE properties are conveniently analyzed with reference to the 
direction of the particle or of the jet with largest $p_T$.
This ``leading" object is expected to reflect the direction of the 
parton produced with the highest transverse momentum in the hard interaction.
Three distinct topological regions in the hadronic final state are thus defined
by the azimuthal angle difference $\Delta\phi$ between the directions,
in the plane transverse to the beam, of the leading object and that of any 
charged hadron in the event.
Hadron production in the ``toward" region with $|\Delta\phi| \!<\! 60^\circ$ and in
the ``away" region with $|\Delta\phi|\!>\! 120^\circ$ is expected to be
dominated by the hard parton-parton scattering and radiation.
The UE structure can be best studied in the ``transverse"
region with $60^\circ \!<\! |\Delta\phi|  \!<\! 120^\circ$. 

UE dynamics is studied through the confrontation of models with the data.
In this paper, MC predictions for charged particle production are compared 
after full detector simulation to the data, uncorrected for 
detector effects.
The predictions for inelastic events are calculated using several tunes of the 
PYTHIA programme, version 6.420~\cite{Sjostrand:1986ep,Sjostrand:2006za}, 
which provide different descriptions of the non-diffractive component:
D6T~\cite{Field:2008zz,rdf1}, DW~\cite{rdf1}, Pro-Q20~\cite{Buckley:2009bj}, Perugia-0
(P0)~\cite{Skands:2009zm}, and CW, the last being adapted from the 
DW tune as described below.
They differ, in particular, in the implementation of the regularization of the
formal $1/\hat{p}_T^4$ divergence of the 
leading order partonic scattering amplitude as the final state parton 
transverse momentum $\hat{p}_T$ approaches $0$.
In PYTHIA this divergence is regularized through the replacement 
$1/\hat{p}_T^4 \rightarrow 1/(\hat{p}_T^2+\hat{p}_{T_0}^2)^2$.
The energy dependence of the cutoff transverse momentum
$\hat{p}_{T_0}$  is parameterized as
$\hat{p}_{T_0}(\sqrt{s})=\hat{p}_{T_0}(\sqrt{s_0}) \cdot (\sqrt{s} \  / \sqrt{s_0})^\epsilon$,
where $\sqrt{s_0}$ is the reference energy at which $\hat{p}_{T_0}$ is determined
and $\epsilon$ is a parameter describing the energy dependence.
CDF studies~\cite{Affolder:2001xt,Acosta:2004wqa} favour a value of
$\hat{p}_{T_0} = 2.0 \GeVc$ for $\sqrt{s_0} = 1.8 \TeV$.
Because a single value of $\hat{p}_{T_0}$ is used to regularize both MPI and hard 
scattering, this parameter governs the description of the amount of MPI 
in the event.
More MPI activity is predicted for smaller values of $\hat{p}_{T_0}$.
 
All tunes considered in this paper are consistent with the UE measurements 
by CDF.
Tunes DW, P0, and Pro-Q20 use $\epsilon = 0.25$, in agreement with CDF data 
at $\sqrt{s} = 630 \GeV$ and $1.8 \TeV$.  
Tune D6T uses the value $\epsilon = 0.16$, which is motivated by the energy 
dependence of charged particle multiplicities measured by the UA5 
collaboration at the ${\rm Sp \bar p S}$ collider~\cite{Moraes:2007rq}.
For tune CW, $\hat{p}_{T_0}$ is decreased to $1.8 \GeVc$ and $\epsilon$ 
is increased to 0.30,  while the parameters controlling the relative weighting of 
possible color connections in the matrix elements are changed back from the DW
values to the PYTHIA defaults;
these changes lead to a large increase of the simulated MPI activity at $\sqrt{s} = 0.9 \TeV$ 
and to an increase of a few percent at the Tevatron with $\sqrt{s} = 1.8 \TeV$, 
while remaining consistent with the CDF results.
The parton distribution functions used to describe the interacting protons are
the CTEQ6LL set for D6T and the CTEQ5L set for the other tunes~\cite{Lai:1999wy,Pumplin:2002vw}.
Tunes P0 and Pro-Q20 use LEP results to describe hadron fragmentation 
at high $z$, where $z$ denotes the fraction of the parton momentum carried by a final 
state particle.
Tune P0 uses the new PYTHIA MPI model~\cite{Skands:2007zg}, which is interleaved with 
parton showering.

\section{ Detector Description and Event Selection}
\label{sec:Detector}

A detailed description of the CMS detector can be found in~\cite{JINST};
features most relevant for the present analysis are described in the following.
A right-handed coordinate system is used with the origin at the nominal interaction 
point (IP). 
The $x$ axis points to the centre of the LHC ring, the $y$ axis is vertical and points upward, 
and the $z$ axis is parallel to the anti-clockwise beam direction. 
The azimuthal angle $\phi$ is measured with respect to the $x$ axis in the $xy$
plane and the polar angle $\theta$ is defined with respect to the $z$ axis.

The pixel and silicon strip tracker, immersed in the axial $3.8\unit{T}$ magnetic field 
provided by the $6  \unit {m}$ diameter superconducting solenoid,
measures charged particle trajectories in the pseudorapidity range 
$|\eta| \!<\! 2.5$, where $\eta = -\ln (\tan(\theta / 2)  )$.
The $p_T$ resolution for $1 \GeVc$ charged particles is between
$0.7\%$ at $\eta = 0$ and $2\%$ at $|\eta| = 2.5$~\cite{JINST}.
The modules composing the tracker system were aligned with cosmic ray 
data taken prior to LHC commissioning, with a precision of 3--4\micron 
in the barrel region~\cite{trackerAlign}.

Three subsystems were involved in the trigger of the detector readout: 
the forward hadron calorimeter (HF),  
the Beam Scintillator Counters (BSC)~\cite{JINST,bsc2}, 
and the Beam Pick-up Timing for eXperiments (BPTX)~\cite{JINST,Aumeyr:2008mt}. 
The steel--quartz-fibre HF covers the region $2.9 \!<\! |\eta| \!<\! 5.2$.  
The two BSCs, each of which consists of a set of 16 scintillator tiles, are located 
along the beam line on each side of the IP at a distance of $10.86 \unit {m}$  
and are sensitive in the range $ 3.23 \!<\! |\eta| \!<\! 4.65$;
they provide information on hits and coincidence signals with an average detection 
efficiency of $96.3\%$ for minimum ionizing particles and a time resolution 
of $3 \unit {ns}$, compared to a minimum inter-bunch spacing of  $25 \unit {ns}$.  
The two BPTX devices, which are located around the beam pipe at a distance of 
$175  \unit {m}$ from the IP, are designed to provide precise information on the structure 
and timing of the LHC beams, with a time resolution better than $0.2 \unit {m}$.
The data analyzed in this paper were selected by requiring a signal in 
both BSC counters, in coincidence with BPTX signals from both beams.
During data taking, interaction rates were typically $11 \unit {Hz}$ and the probability for 
multiple inelastic collisions to occur in the same proton bunch crossing was less 
than $2 \times 10^{-4}$.  

\begin{table}
\caption   
{
\label{tab:selectEvents}
Numbers of events in the data satisfying the selection criteria, and corresponding cumulative event 
fractions in the data and for the simulation based on PYTHIA with tune D6T.
In the lower part of the table, the effects of various selection cuts applied to the leading 
object with \etacut\ are given, each fraction being given with respect to the previous cut.
}
\begin{center}
\begin{tabular}{r|c|c|c} 
Event selection       & Data [nb. events] & Data [\%] & MC [\%] \\
\hline
\hline
triggered                                      	          	& 255 122        &  100     &   100     \\
+ 1 primary vertex                                    	 & 239 038       &  93.7    &   92.9    \\
+ 15 cm vertex $z$ window              	& 238 977        &  93.7    &   92.8   \\
+ at least 3 tracks associated                                   & 230 611        &  90.4    &   88.7   \\
\hline\hline
leading track, $p_{T}\!>\!0.5 \GeVc$                  &  216 215     &  93.8    &   93.2   \\
	       $p_{T}\!>\!1.0 \GeVc$                  &  131 421     &  60.8   &   55.0   \\
                          $p_{T}\!>\!2.0 \GeVc$                  &  \,28 210        & 21.5    &   19.5   \\
\hline
leading track-jet, $p_{T}\!>\!1.0 \GeVc$            &    155 005    &  67.2   &   62.9   \\
                               $p_{T}\!>\!3.0 \GeVc$            &    \,24 928      &  16.1   &  15.9    \\
\end{tabular}
\end{center}
\end{table}

The event selection requires one reconstructed primary vertex~\cite{CMS-PAS-TRK-10-001}
with $z$ coordinate within $15  \unit {cm}$ 
of the centre of the beam collision region, of which the rms size is 
about $4  \unit{cm}$.
Three or more tracks must be identified as originating at the vertex.
Table~\ref{tab:selectEvents} gives the numbers of events that pass 
these selection criteria.
A study of data collected with non-colliding beams showed that beam-induced 
backgrounds are negligible.

Kinematic selections are based on the transverse 
momentum of the leading charged particle or of the leading track-jet,
which must be reconstructed with pseudorapidity \etacut.
The leading charged particle, or ``leading track", must be reconstructed in the 
tracking detector.
The leading track-jet is defined using the SISCone algorithm~\cite{Salam:2007xv} 
as implemented in the fastjet package~\cite{Cacciari:2005hq} with a clustering 
radius $R = \sqrt {(\Delta \phi)^2 + (\Delta \eta)^2 } = 0.5$.
Only charged particles reconstructed in the tracker, with \ptcut\ and 
$|\eta|\!<\!2.5$, are used to define the track-jet.
No further correction is applied to the 
track-jet $p_T$.
The $\eta$~range of the charged particles used to define the track-jet ($|\eta|\!<\!2.5$) is 
chosen to be wider than that used for the UE analysis (\etacut ) 
in order to avoid a kinematic bias.
A simulation-based study of jets with $p_T \!>\! 5 \GeVc$ indicates that track-jets in CMS 
are found with high efficiency and good angular and energy 
resolutions~\cite{pasjme08001}; 
this has been verified for softer jets in the present analysis.
The results of selection cuts on the leading track and leading track-jet $p_T$ 
are given in Table~\ref{tab:selectEvents}.

A detailed simulation of the CMS detector response was performed, based on
the GEANT4 package~\cite{Agostinelli:2002hh} with event simulation using PYTHIA 
tune D6T.
The position and shape of the beam interaction region were adjusted to agree
with the data~\cite{CMS-PAS-TRK-10-001}. 
Simulated events were processed and reconstructed in the same manner 
as collision data, and 
the results of the simulation are also reported in Table~\ref{tab:selectEvents}.

\section{Track Selection and Systematic Uncertainties}
\label{sec:Track}

\begin{table}
\caption{
\label{tab:selectTracks}
Numbers of tracks in the selected event sample 
for successive track selection criteria, and corresponding fractions 
in the data and for the simulation based on PYTHIA with tune D6T.
Each fraction is given with respect to the previous selection cut.
}
\begin{center}
\begin{tabular}{r|c|c|c} 
 Track selection          & Data [nb. tracks]	& Data [\%]         	& MC [\%] \\ 
 \hline
\hline  
reconstruction algorithm           & 4 004 923                 & 100  & 100  \\ 
+ $p_T \!>\! 0.5 \GeVc$                   & 1 707 998                 & 42.6 & 44.0 \\ 
+ $|\eta|\!<\!2.5$                               & 1 689 910                 & 98.9 & 98.7  \\ 
+ $|\eta|\!<\!2$                                   & 1 399 344                 & 82.8 & 81.5  \\ 
+ $d_{xy}/\sigma(d_{xy})\!<\!5$      & 1 235 193                 & 88.3 & 88.8  \\ 
+ $d_z/\sigma(d_z)\!<\!5$               & 1 204 979                 & 97.6 & 97.9 \\ 
+ $\sigma (p_T) / p_T \!<\! 5\%$    & 1 168 530                 & 97.0 & 96.9 \\ 
\hline \hline  
Total                                                 & 1 168 530                 & 29.2 & 29.8 \\
\end{tabular}
\end{center}
\end{table}

A charged particle track is selected for the UE analysis if it 
originates from the primary vertex and is reconstructed
in the pixel and silicon strip 
tracker with transverse momentum \ptcut\ and pseudorapidity \etacut.
A high purity reconstruction algorithm (see Section~3 of~\cite{CMS-PAS-TRK-10-001})
is used, which keeps low levels of fake and poorly reconstructed tracks.
To decrease contamination by secondary tracks from decays of long-lived 
particles and photon conversions, 
the distance of closest approach between track and primary vertex is required to be 
less than five times its estimated uncertainty,
both in the transverse plane, $d_{xy} / \sigma(d_{xy}) \!<\! 5$, 
and along the $z$ axis, $d_z/\sigma(d_z) \!<\! 5$.
Poorly measured tracks are removed by requiring 
$\sigma(p_T)/p_T \!<\! 5\%$, where $\sigma(p_T)$ is the uncertainty on the 
transverse momentum measurement. 
In the selected track sample with \etacut, these cuts result in a background
level of  3\%, 1\% from $K_{S}^0$ and  $\Lambda^0$ decay products 
and 2\% from fake tracks.

The numbers of tracks accepted at the different selection steps and 
the corresponding fractions are given in Table~\ref{tab:selectTracks},
together with the fractions calculated using simulated data.
Agreement is observed at the percent level between data and simulation,
for all selection steps.

\begin{table}
\caption{\label{tab:systematics-leadjet}
Systematic uncertainties on track selection and reconstruction (see description
in text).
The uncertainties, expressed in \%, are quoted for characteristic values 
of variables used for UE studies in the transverse region. For the first two variables,
$p_T$ designates the minimal value of the track-jet $p_T$;
for the last three variables, events with a leading track-jet with $p_T \!>\! 3 \GeVc$
are selected.
}
\begin{center}
\begin{tabular}{c|c|c|c|c|c|c|c|c}

       & {\small track} & {\small tracker} & {\small tracker} & {\small bg.}   & {\small trigger} & {\small dead} & {\small beam} & {\small total}   \\
       & {\small sel.}    & {\small align.}  & {\small mater.}  & {\small cont.}  &                            & {\small ch.}      & {\small spot}    &                            \\
\hline
\hline
${\rm d}^{2}N_{\rm {ch} } / {\rm d}\eta {\rm d(}\Delta \phi {\rm )}$ ($p_T = 3.5 \GeVc$)                       
       &  0.3   &  0.3  &  1.0  &  0.8  &  0.6  &  0.1  &  0.5  & 1.8 \\
${\rm d}^{2}\Sigma p_{T} / {\rm d}\eta {\rm d(}\Delta \phi{\rm )}$ ($p_T = 3.5 \GeVc$)             
       &  0.4  &  0.3   &  1.0 &  0.8  &  1.1   &  0.1  &  0.5  & 1.8 \\
${\rm d}N_{\rm {ev} } / {\rm d}N_{\rm {ch} }$ ($N_{\rm {ch} } = 4$)                                                                  
       &  0.6  &  0.6   &  1.2  &  1.0  &  1.2  &  0.2  &  0.6  & 2.3 \\
${\rm d}N_{\rm {ev} } / {\rm d}\Sigma p_{T}$ ($\Sigma p_{T} = 4.5 \GeVc$)                        
       &  0.5  &  0.2   &  0.6  &  0.5  &  1.2  &  0.2  &  0.4  & 1.6 \\
${\rm d}N_{\rm {ch} } / {\rm d}p_{T}$ ($p_{T} = 1 \GeVc$)                                                       
       &  0.8  &  0.6   &  1.0  &  0.8  &  1.0  &  0.2  &  0.5  & 2.0 \\

\end{tabular}
\end{center}
\end{table}

Several systematic uncertainties may affect the comparison of models 
with the data.
The sources of these uncertainties
include the implementation in the simulation of track selection criteria, tracker alignment and 
tracker material content, background contamination, trigger conditions, and run-to-run 
variations of tracker and beam conditions.

The uncertainty in the simulation of track selection 
has been evaluated by applying various sets of criteria and comparing
their effects to the data and to simulated events.

The tracking performance depends on occupancy;
because efficiencies and fake rates computed using different models are found 
to be consistent within statistical uncertainties, no systematic uncertainty 
 due to occupancy variation is assigned.
The effects of  tracker misalignment are found to change the results by less than 1\%.  
The description in the simulation of inactive tracker material has been found to be 
adequate within 5\%; 
increasing the material densities by 5\% in the simulation induces a change 
smaller than 1\% in the tracking efficiency and has no significant effect on 
background rates. 

The simulation has been found to underestimate $K_{S}^0$ and $\Lambda^0$ production 
as well as photon conversion rates.
These discrepancies induce changes of less than 0.5\% in the background 
contamination.
Increasing the combinatorial background by a conservative 
30\% leads to a combined 0.8\% uncertainty due to background description.

The uncertainty related to the simulation of the BSC-based trigger is taken
to be half of the difference between the distributions obtained with and without 
trigger simulation.
This estimate of the trigger-related systematic uncertainty was
verified by means of HF-triggered events for which the BSCs had not 
generated a trigger.

The number of inactive tracker channels changes from run to run; 
reproducing this effect in the simulation induces a change of less than 0.5\%
in the observed distributions. 
The beam collision region is not perfectly centred within the detector, and its 
position changes from run to run;
simulating different beam spot positions, consistent with those observed in
different runs, leads to a  0.5\% uncertainty.

The systematic uncertainties are largely independent from one another, but they are 
correlated among data points in the experimental distributions. 
Table~\ref{tab:systematics-leadjet} gives the main uncertainties for selected events 
with a leading track-jet with $p_T \!>\! 3 \GeVc$,
for characteristic values of variables used for UE studies in the transverse region.
Most uncertainties increase by typically 50\% when the selection requires a
leading track with $p_T \!>\! 2 \GeVc$.

\section{Results}
\label{sec:Results}

Predictions from the various PYTHIA models, after full detector simulation, are compared to
the data.
The scale of an interaction at parton level is defined by the $p_T$ value 
of the leading object, either a track or a track-jet with \etacut.
As can be observed in Table~\ref{tab:selectEvents}, demanding a leading particle with 
$p_T \!>\! 2 \GeVc$ or a leading track-jet with $p_T \!>\! 3 \GeVc$ reduces the sample
size by a similar factor of about~$10$.

\begin{figure}[htbp]
\begin{center}
           \includegraphics[angle=90,width=0.45 \textwidth]{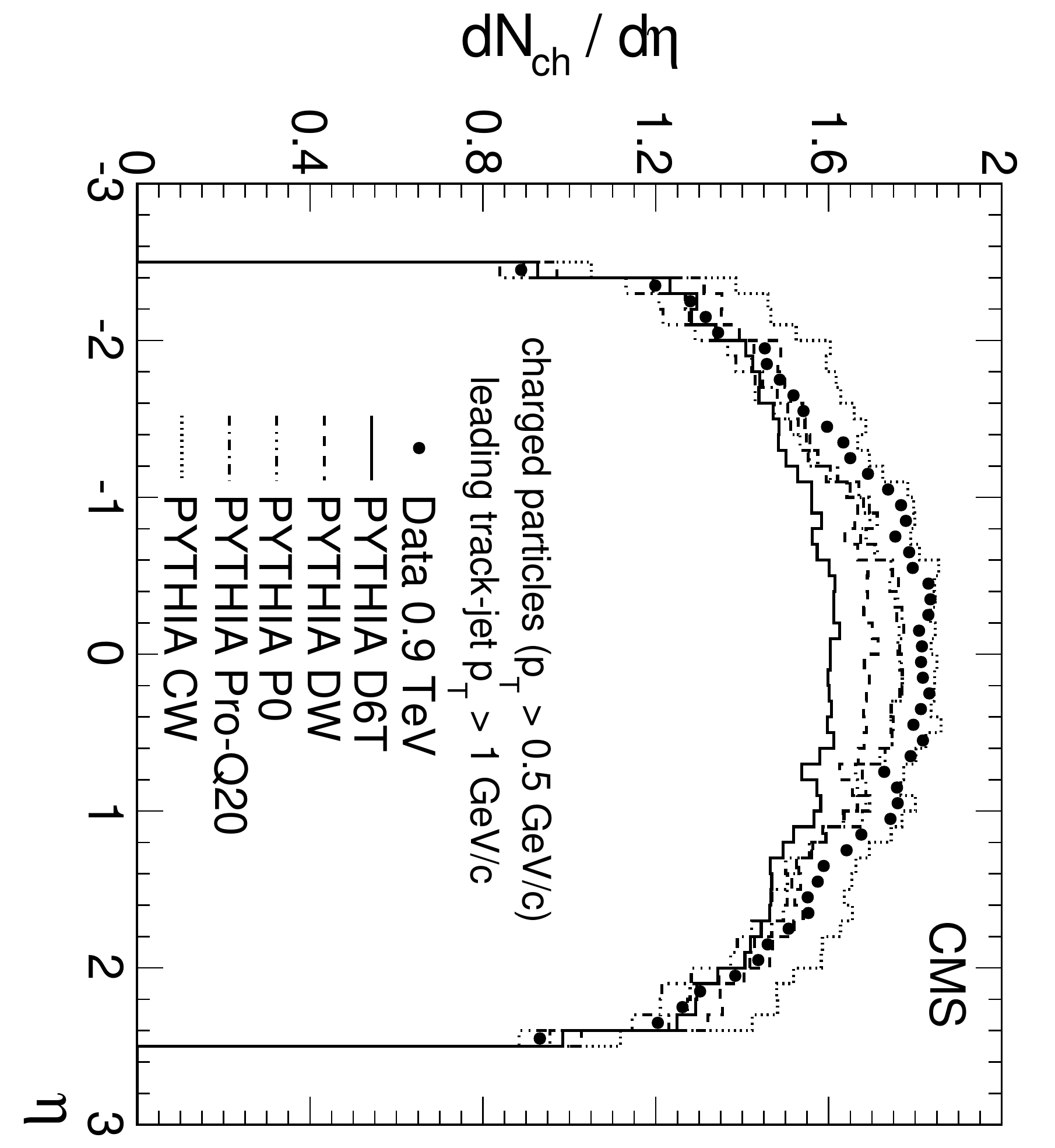}
           \includegraphics[angle=90,width=0.45 \textwidth]{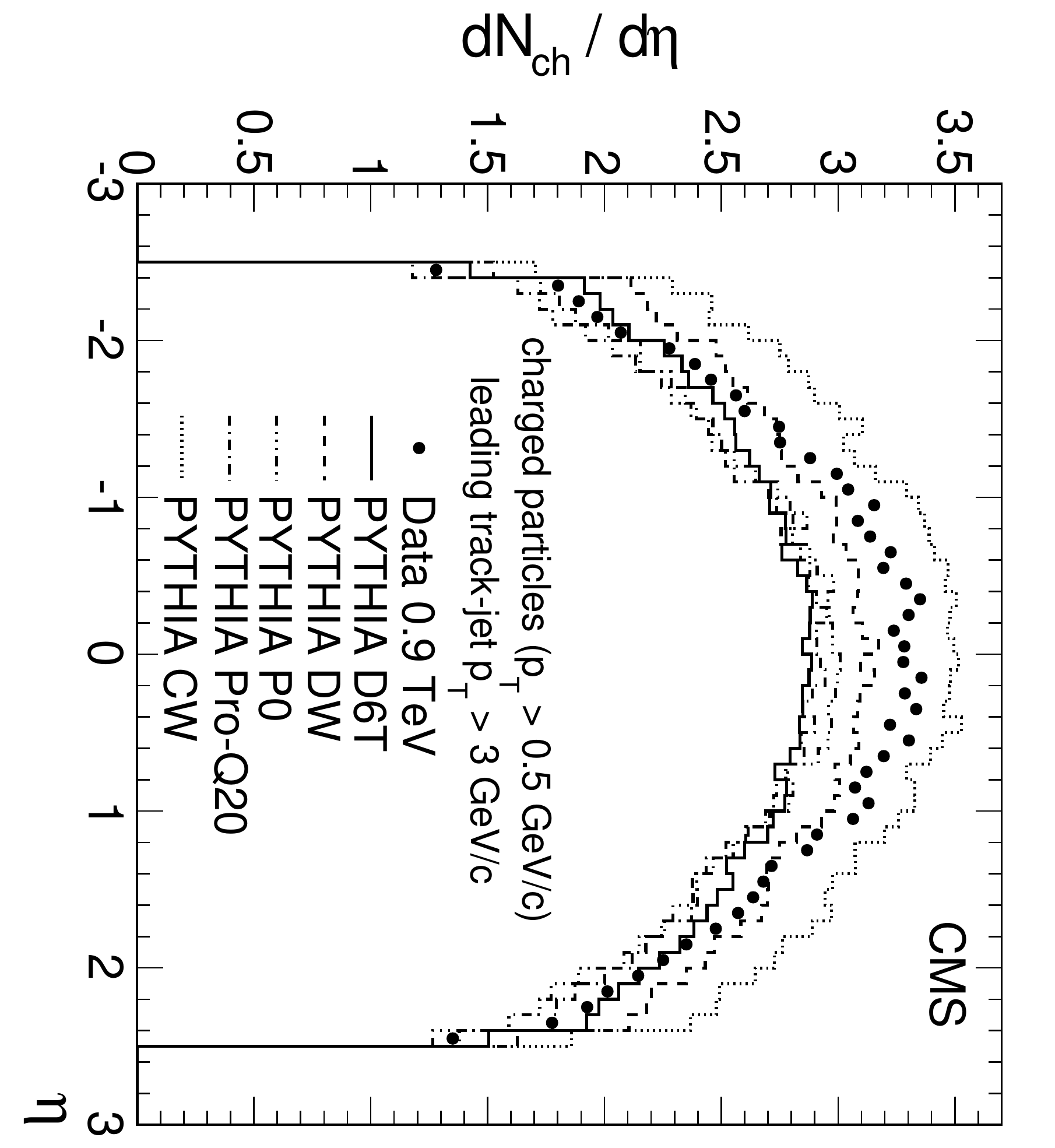}
\caption{
Average multiplicity, per unit of pseudorapidity, of charged particles with \ptcut, as a function 
of $\eta$.
The leading track-jet is required to have $|\eta| \!<\! 2$ and 
(left) $p_T \!>\! 1 \GeVc$, or  (right) $p_T \!>\! 3 \GeVc$ 
(note the different vertical scales).
Predictions from several PYTHIA MC tunes, including full detector simulation, are compared 
to the data. 
\label{fig:etascale}}
\end{center}
\end{figure}

Figure~\ref{fig:etascale} presents, as a function of $\eta$,
the average multiplicity $N_{\rm {ch} }$ per unit of pseudorapidity of charged particles
with \ptcut;
for this figure, the track selection is extended to $|\eta| = 2.5$.
Distributions are shown for two choices of the minimal value required for
the $p_T$ of the leading  track-jet.
For a harder scale, the multiplicities are larger and charged particles with \ptcut\
are produced more centrally.
The various PYTHIA tunes describe several features of the data: 
the overall normalization, 
the $\eta$ dependence of particle production,
and the effect of the leading track-jet $p_T$ cut.
However, no simulation describes perfectly all elements of the data, either in 
normalization or in shape. 
For both values of the minimal  $p_T$ of track-jets, the data show a significantly stronger 
$\eta$ dependence than predicted by the PYTHIA tunes.
Predictions of tune CW are too high in normalization,  whereas those of 
tunes D6T, P0, and Pro-Q20 are generally too low, with DW being too low in the central 
region and too high at large $|\eta|$ values.
The shape description is slightly better with tunes P0 and Pro-Q20.
Similar observations are made when the selection criteria are applied to the leading 
track $p_T$.
The observed features can be due to shortcomings in the description of parton
fragmentation and radiation (essentially the toward and away regions), in the description of
the UE (visible in the transverse region), or in both.

\begin{figure}[htbp]
\begin{center}
\includegraphics[angle=90, width=0.45 \textwidth]{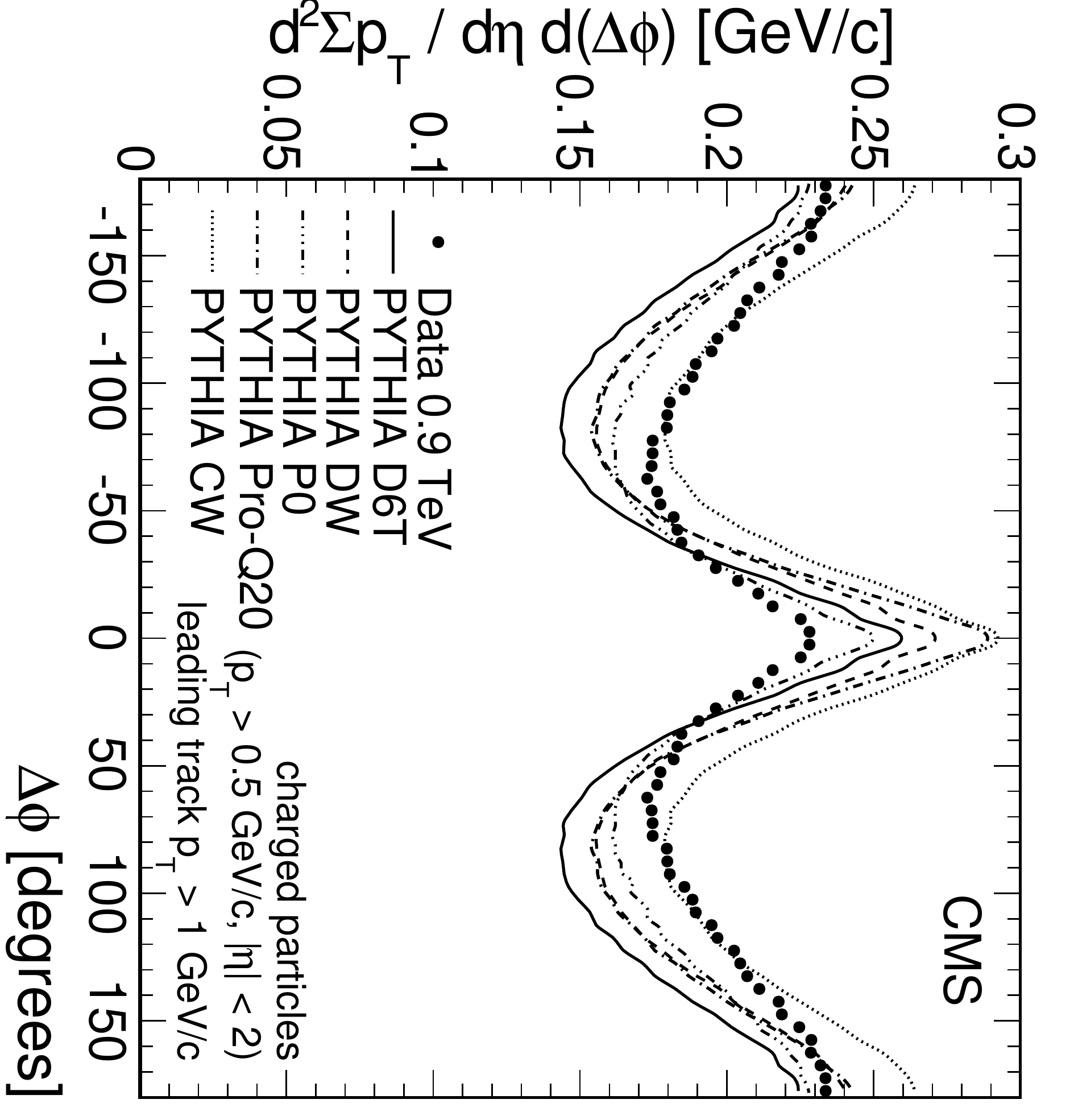}
\includegraphics[angle=90, width=0.45 \textwidth]{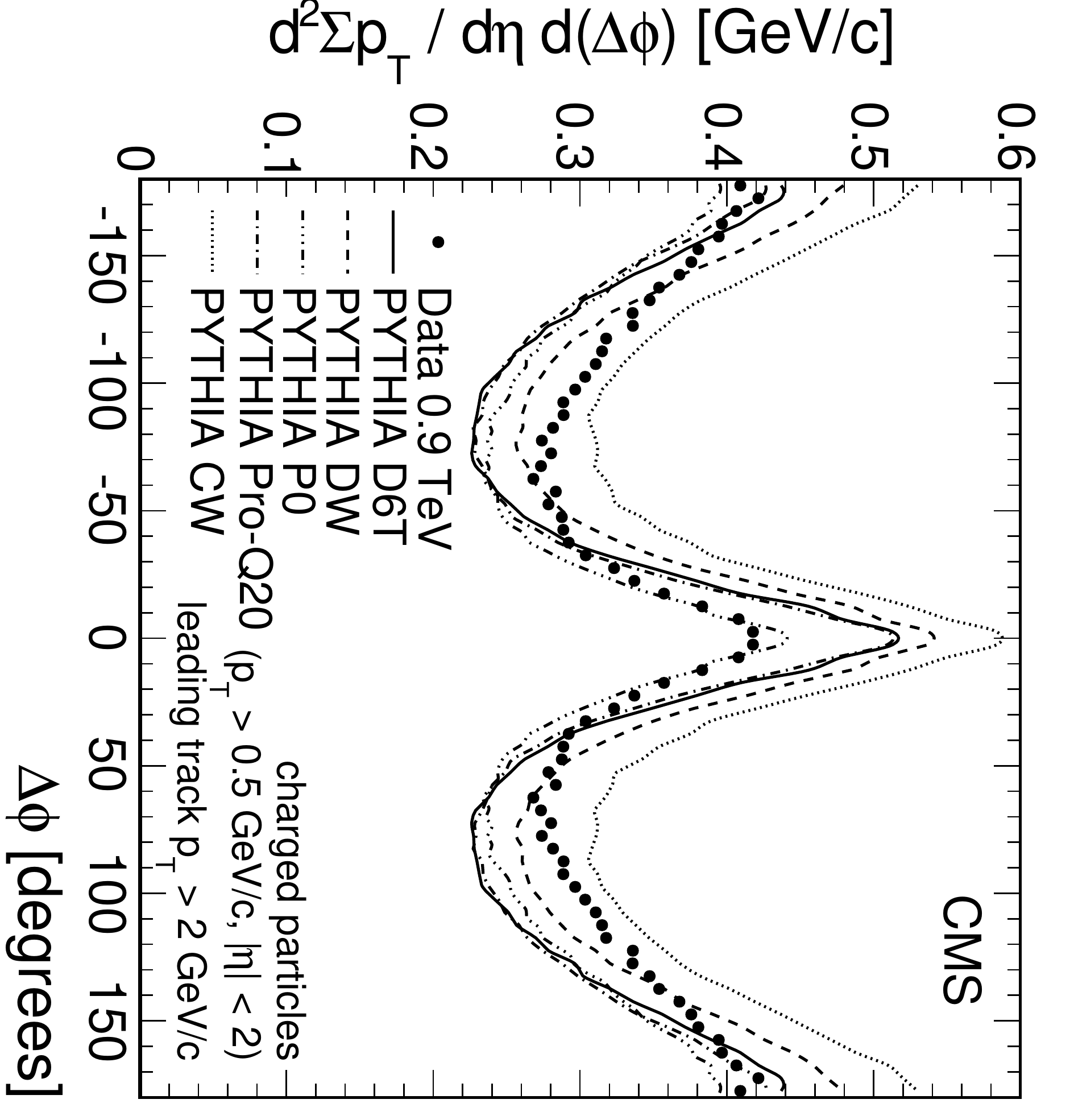}
\end{center}
\caption{
Average scalar sum of transverse momenta of charged particles with \ptcut\ and 
\etacut, per unit of pseudorapidity  and per radian, 
plotted as a function of the azimuthal angle difference $\Delta \phi$ 
relative to the leading track
(the measurements have been symmetrized in $\Delta \phi$).
The leading track, which is excluded from the $p_T$ sum, is required to have $|\eta| \!<\! 2$ and 
(left)~$p_T \!>\! 1 \GeVc$, or (right)~$p_T \!>\! 2 \GeVc$
(note the different vertical scales).
Predictions from several PYTHIA MC tunes, including full detector simulation, are compared 
to the data. 
\label{fig:deltaPhi} }
\end{figure}

The production of charged particles with \ptcut\ and \etacut\ in the different 
topological regions and the quality of the MC
description can be examined through the distribution of the azimuthal 
separation \delphi\ between the directions of the leading object and of any 
selected track.
As an example, Fig.~\ref{fig:deltaPhi} presents the distribution of  
${\rm d}^{2}\Sigma p_{T} / {\rm d}\eta {\rm d(}\Delta \phi {\rm )}$, 
where $\sum p_T$ denotes the scalar sum of particle transverse momenta,
excluding the leading track at $\Delta\phi  = 0$.
The events are selected with two different values of the leading track minimal $p_T$. 
The characteristic features of two-jet parton-parton production with underlying
activity are observed.
Although the leading track $p_T$ is not included in the calculation, the 
average $\sum p_T$ in the toward region, $|\Delta\phi| \!<\! 60^\circ$, 
shows substantial activity due to parton fragmentation and radiation.
Charged hadron production is also significant around the opposite direction, 
$|\Delta\phi|\!>\! 120^\circ$;
this is attributed to the fragmentation of the second outgoing parton.
In the transverse region with $60^\circ\!<\! |\Delta\phi| \!<\! 120^\circ$,
hadron production is depleted but it is nonzero, a feature that is attributed mainly 
to MPI.
Similar features of the event structure are observed for the
average track multiplicity
and for selections based on the leading track-jet $p_T$.

In the toward region, 
all PYTHIA tune predictions are significantly above the data, 
except for tune P0 with the scale $p_T \!>\! 2 \GeVc$.
The poor description by tune Pro-Q20 compared to that of P0 may appear
surprising since both use LEP results on jet fragmentation.
A difference between these tunes is that P0 incorporates newer MPI modeling 
and $p_T$ ordered showering.
Model descriptions are better for the away region, except for the CW and DW tunes, 
both of which are significantly above the data when the scale is large.

The transverse region is most relevant for understanding UE properties.
Here, the best tunes are CW and DW.
The predictions of the CW model are slightly too high, especially for the higher $p_T$ 
scale, and those of DW slightly too low;
predictions of the other tunes are even lower.
In the following, studies of the UE using the transverse region 
will focus on the comparison with data of the predictions of the CW and DW tunes.

\begin{figure}[htbp]
\begin{center}
\includegraphics[angle=90, width=0.45 \textwidth]{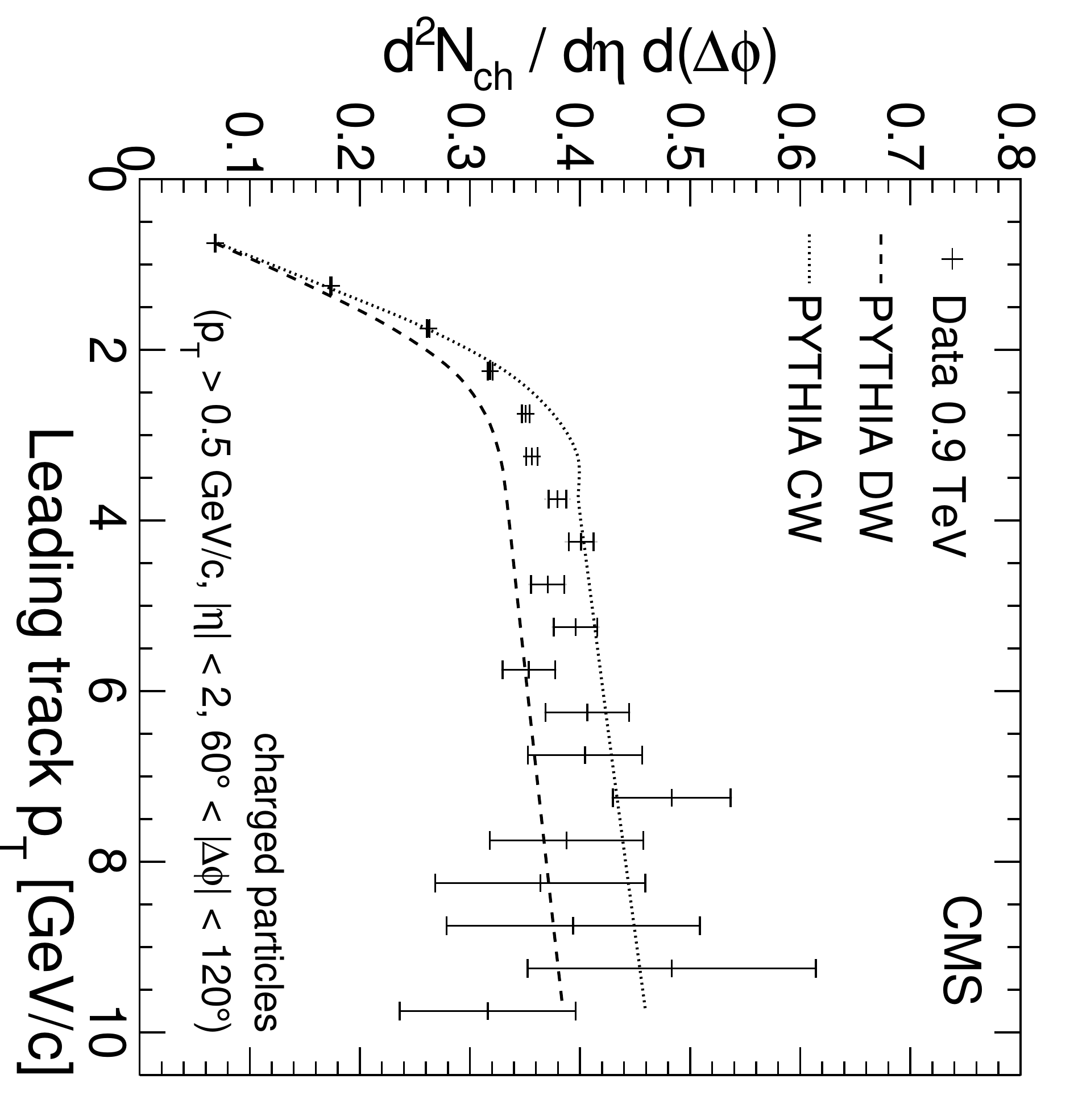}
\includegraphics[angle=90, width=0.45 \textwidth]{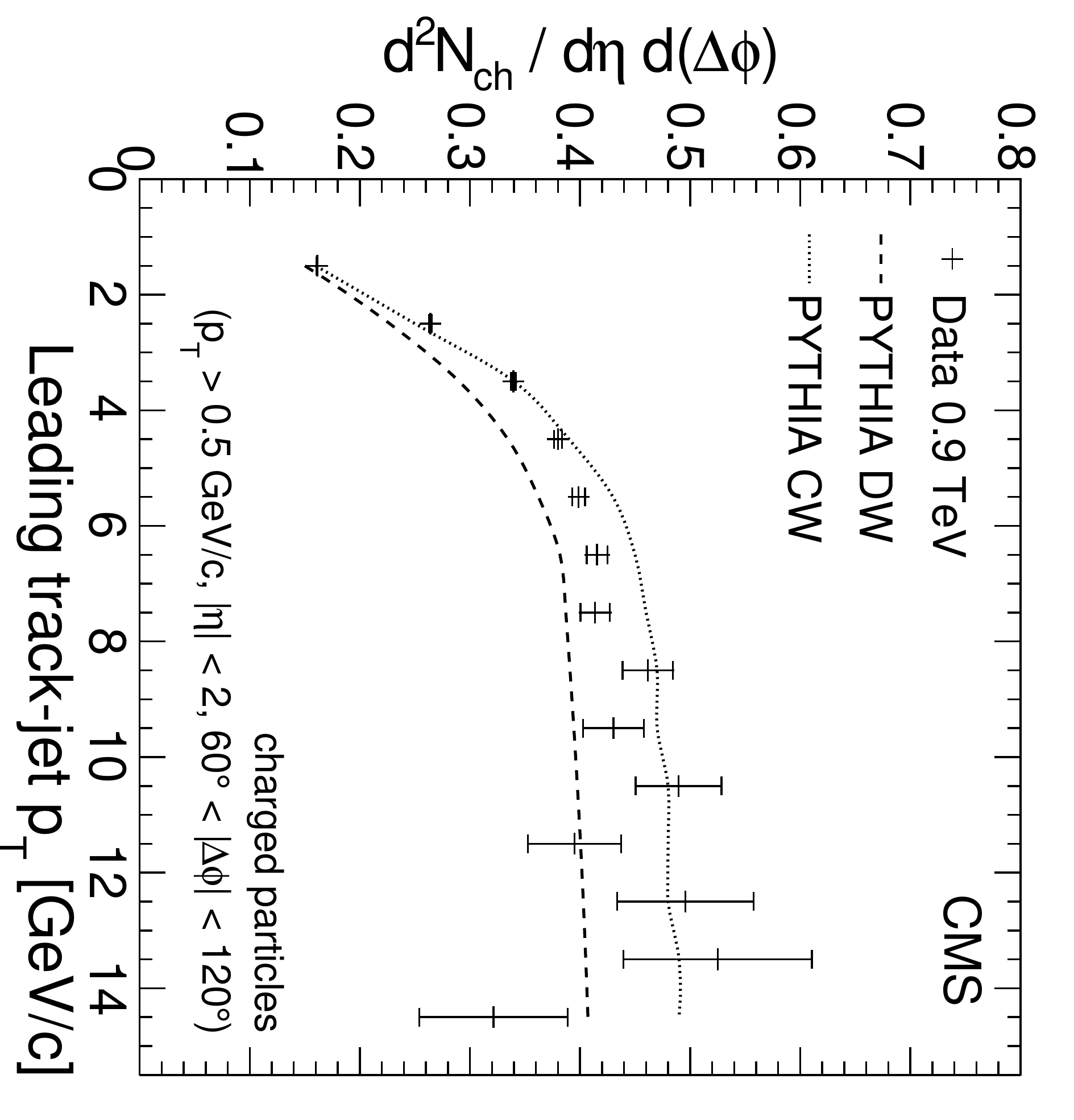} \\
\includegraphics[angle=90, width=0.45 \textwidth]{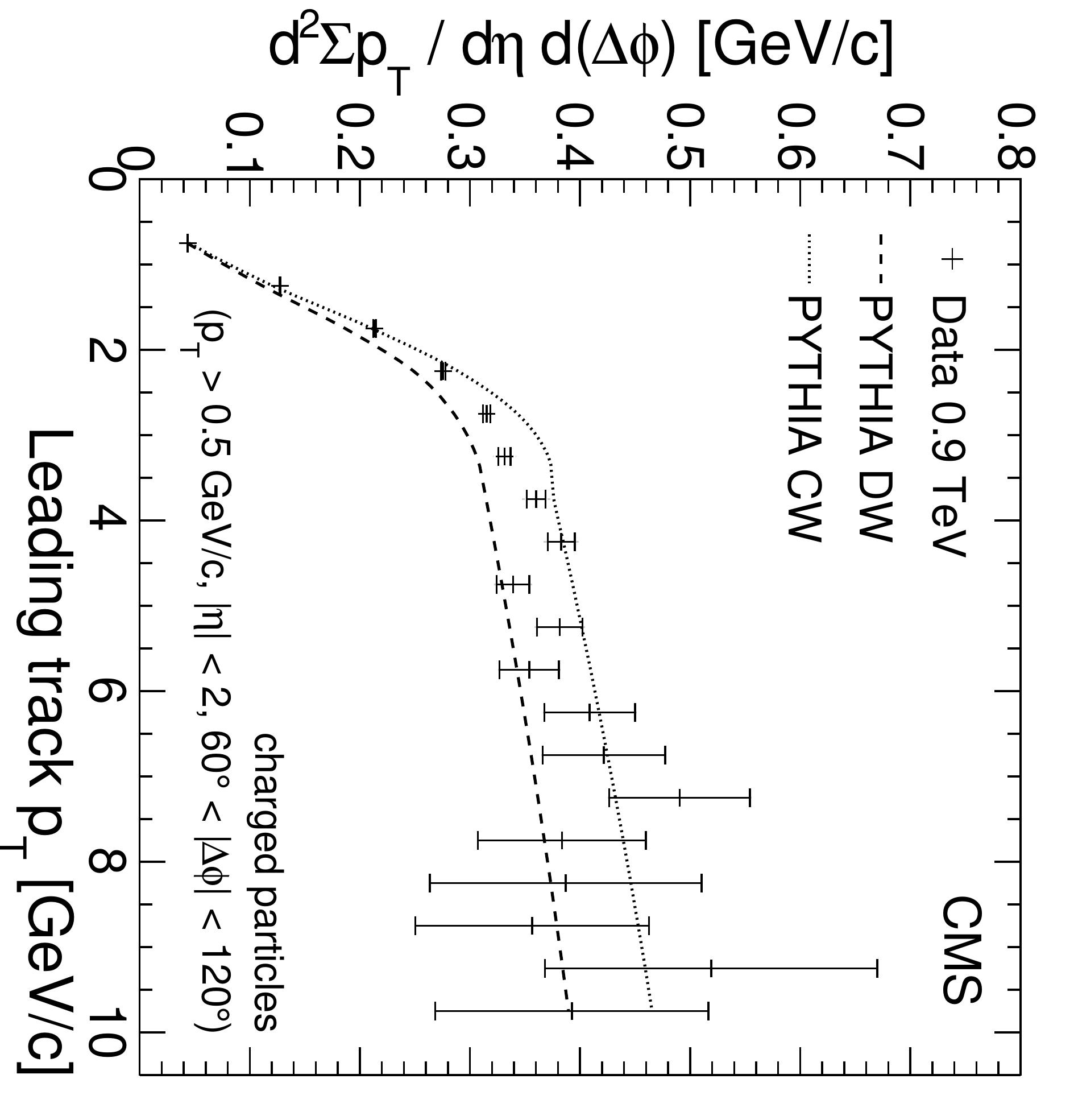}
\includegraphics[angle=90, width=0.45 \textwidth]{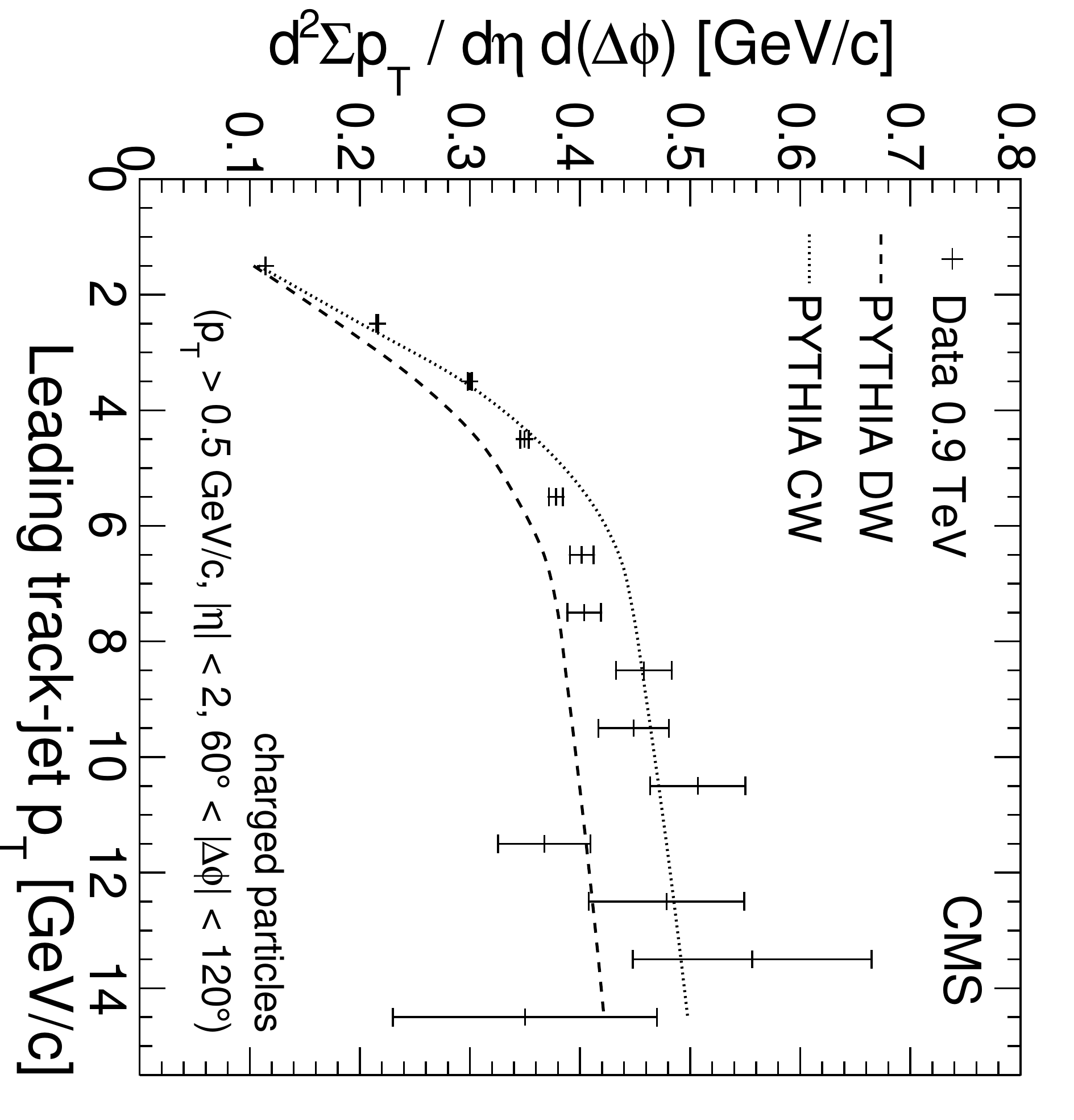}
\end{center}
\caption{ 
For charged particles with \ptcut\ and \etacut\ in the transverse region,
$60^\circ\!<\! |\Delta\phi| \!<\! 120^\circ$:
(upper plots) average multiplicity, 
and 
(lower plots) average scalar $\sum p_T$, per unit of pseudorapidity  and per radian,
as a function of 
(left plots) the $p_T$  of the leading track, and 
(right plots) the $p_T$  of the leading track-jet.
The inner error bars indicate the statistical uncertainty and the outer error bars 
the total experimental uncertainty (statistical and systematic uncertainties
added in quadrature);
statistical errors dominate.
Predictions of the DW and CW \py\ MC tunes, including full detector simulation, are 
compared to the data. 
\label{fig:CWDWToData}}
\end{figure}

\begin{figure}[htbp]
\begin{center}
\includegraphics[angle=90, width=0.45 \textwidth]{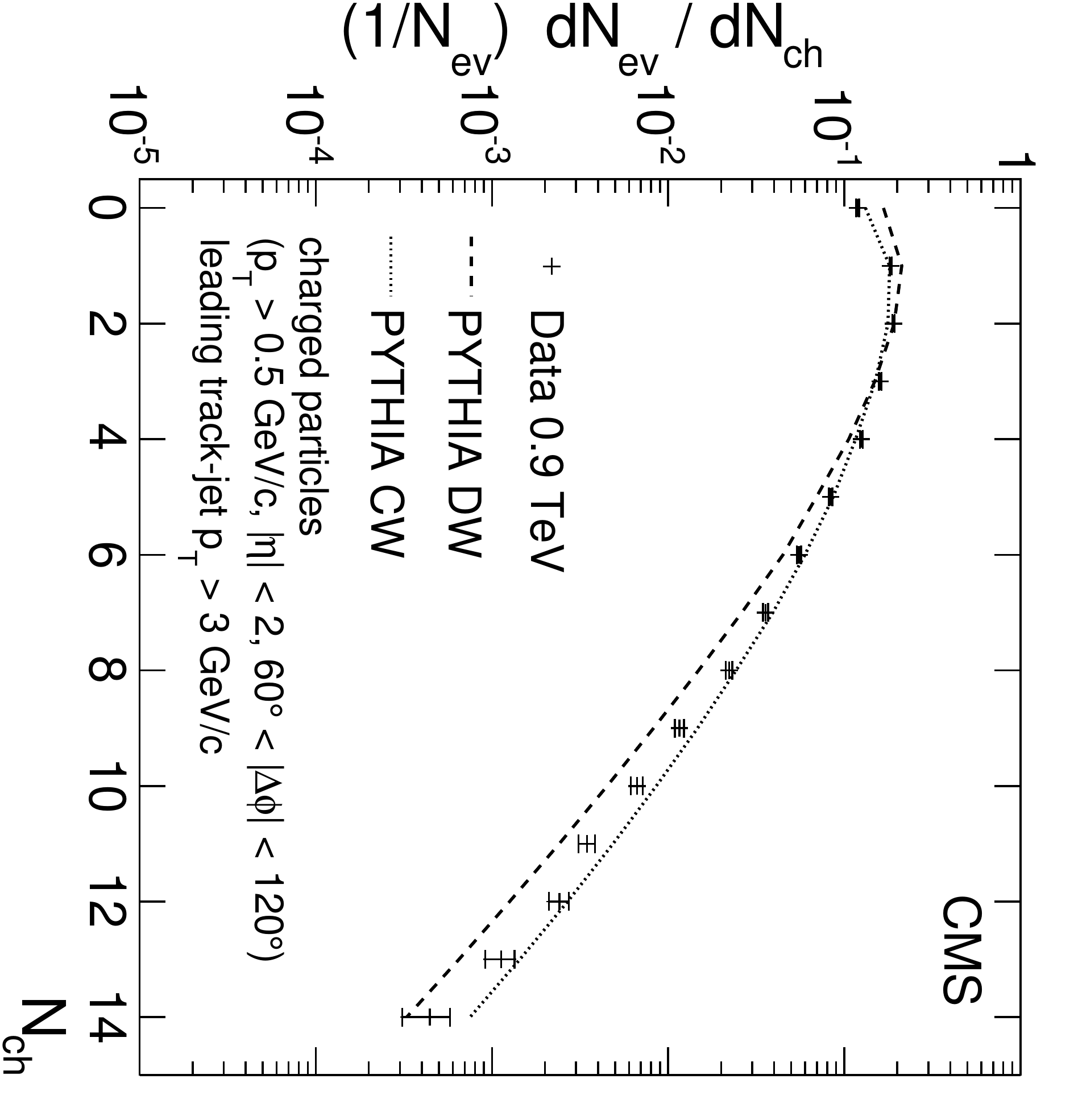}
\includegraphics[angle=90, width=0.45 \textwidth]{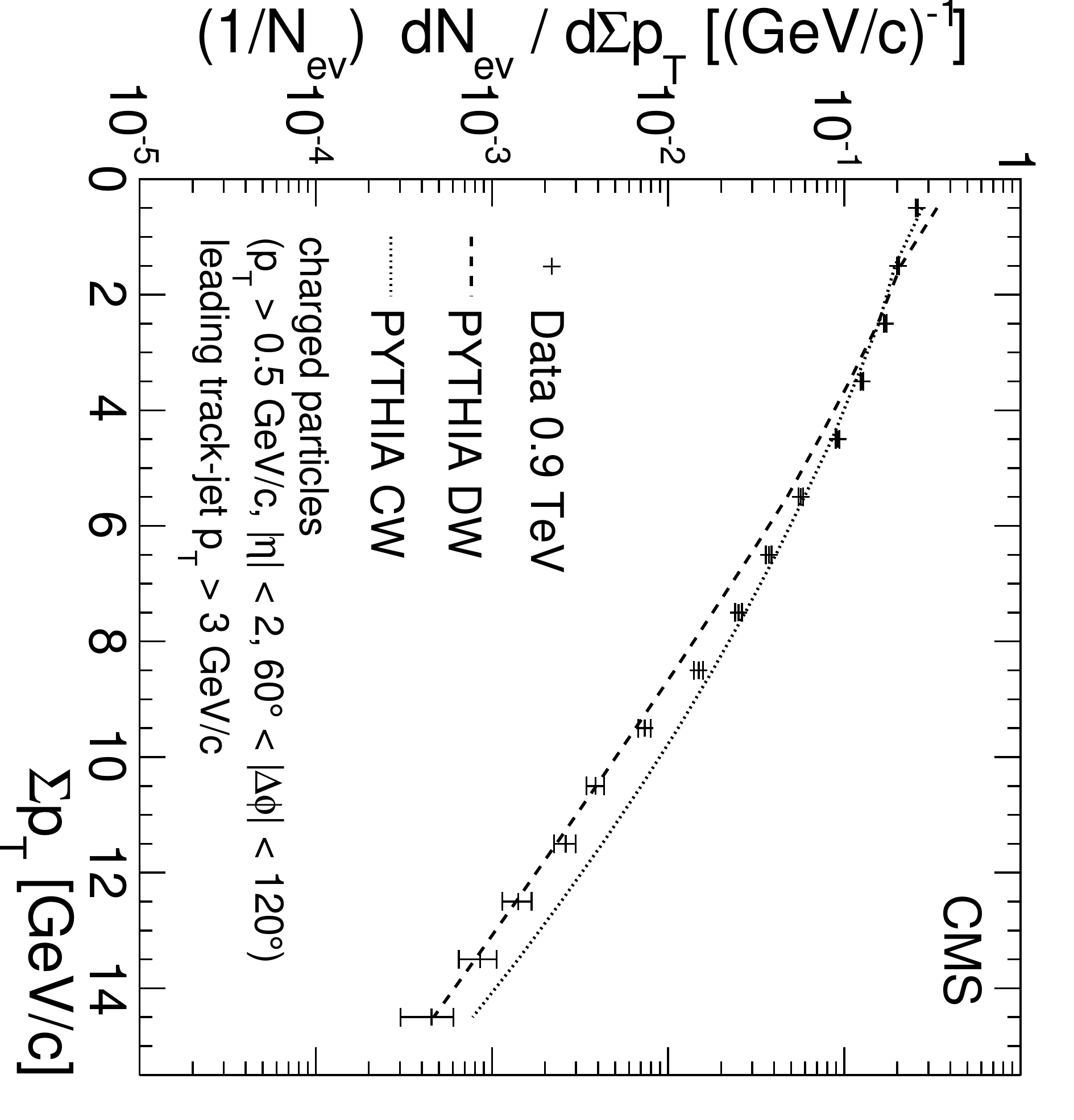} \\
\includegraphics[angle=90, width=0.45 \textwidth]{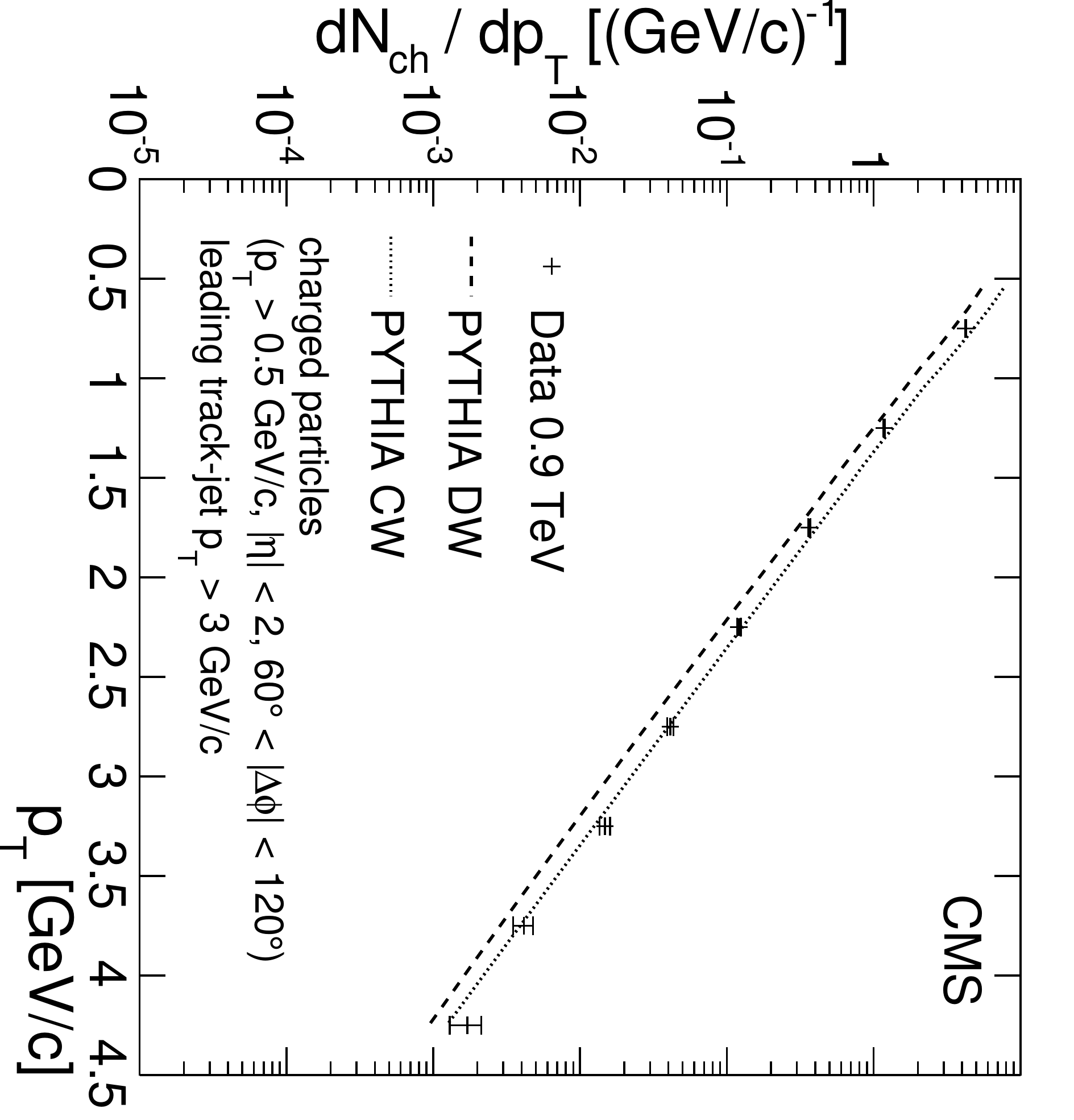}
\end{center}
\caption{
For charged particles with \ptcut\ and \etacut\ in the transverse region,
$60^\circ\!<\! |\Delta\phi| \!<\! 120^\circ$:
(upper left)~normalized multiplicity distribution;
(upper right)~normalized scalar $\sum p_T$ distribution; 
(bottom)~$p_T$ spectrum.
The leading track-jet is required to have $|\eta| \!<\! 2$ and $p_T \!>\! 3 \GeVc$.
The inner error bars indicate the statistical uncertainty and the outer error bars 
the total experimental uncertainty (statistical and systematic uncertainties
added in quadrature);
statistical errors dominate.
Predictions of the DW and CW PYTHIA MC tunes, including full detector simulation, are 
compared to the data. 
\label{fig:diststoDWCW} 
}
\end{figure}

Figures~\ref{fig:CWDWToData} and~\ref{fig:diststoDWCW} provide detailed 
information on the production of charged particles
with \ptcut\ and \etacut\ in the transverse region with 
$60^\circ\!<\! |\Delta\phi| \!<\! 120^\circ$.
Figure~\ref{fig:CWDWToData} presents the distributions of the average 
multiplicity, ${\rm d}^{2}N_{\rm {ch} }/{\rm d}\eta {\rm d(}\Delta\phi{\rm )}$,
and of the average scalar momentum sum, 
${\rm d}^{2}\Sigma p_{T}/{\rm d}\eta {\rm d(}\Delta\phi{\rm )}$,
as a function of the scale provided by the $p_T$ of the leading track 
or of the leading track-jet.
At low $p_T$ of the leading object, the multiplicity and the scalar $\sum p_T$ rise rapidly 
with $p_T$, which is attributed to MPI.
This fast rise is followed by a slower increase for 
leading tracks with $p_T \gsim 3 \GeVc$ (left plots)
or leading track-jets with $p_T \gsim 4 \GeVc$ (right plots),
attributed to a saturation of MPI, plus additional  radiation;
as expected, a similar scale is provided by a lower $p_T$ value for a 
leading track than for a leading track-jet.
The behaviour of the data is reproduced by both the CW and DW tunes, as well 
as by the other PYTHIA tunes (not shown), with a better description by CW in 
the low $p_T$ region.

The distributions of charged particle multiplicity, of scalar $\sum p_T$, 
and of particle $p_T$ are presented in Fig.~\ref{fig:diststoDWCW} for 
events selected with a leading track-jet with $p_T \!>\! 3 \GeVc$. 
The CW and DW tunes bracket the data over most of the
experimental range, and they describe the 
various dependences rather well.
Similar behaviours are observed for selections based on the leading 
track $p_T$.

\begin{figure}[htbp]
\begin{center}
\includegraphics[angle=90, width=0.45 \textwidth]{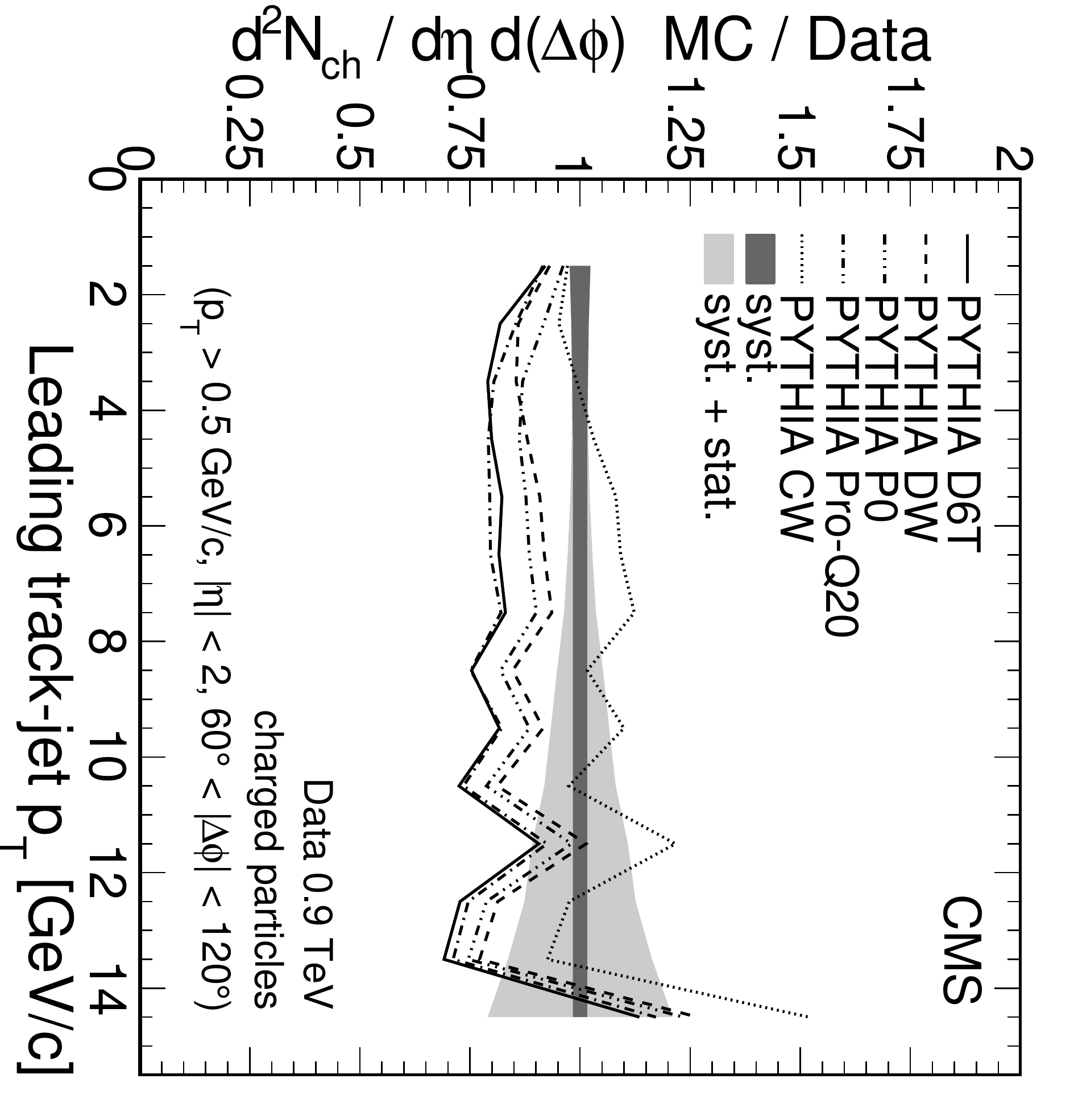}
\includegraphics[angle=90, width=0.45 \textwidth]{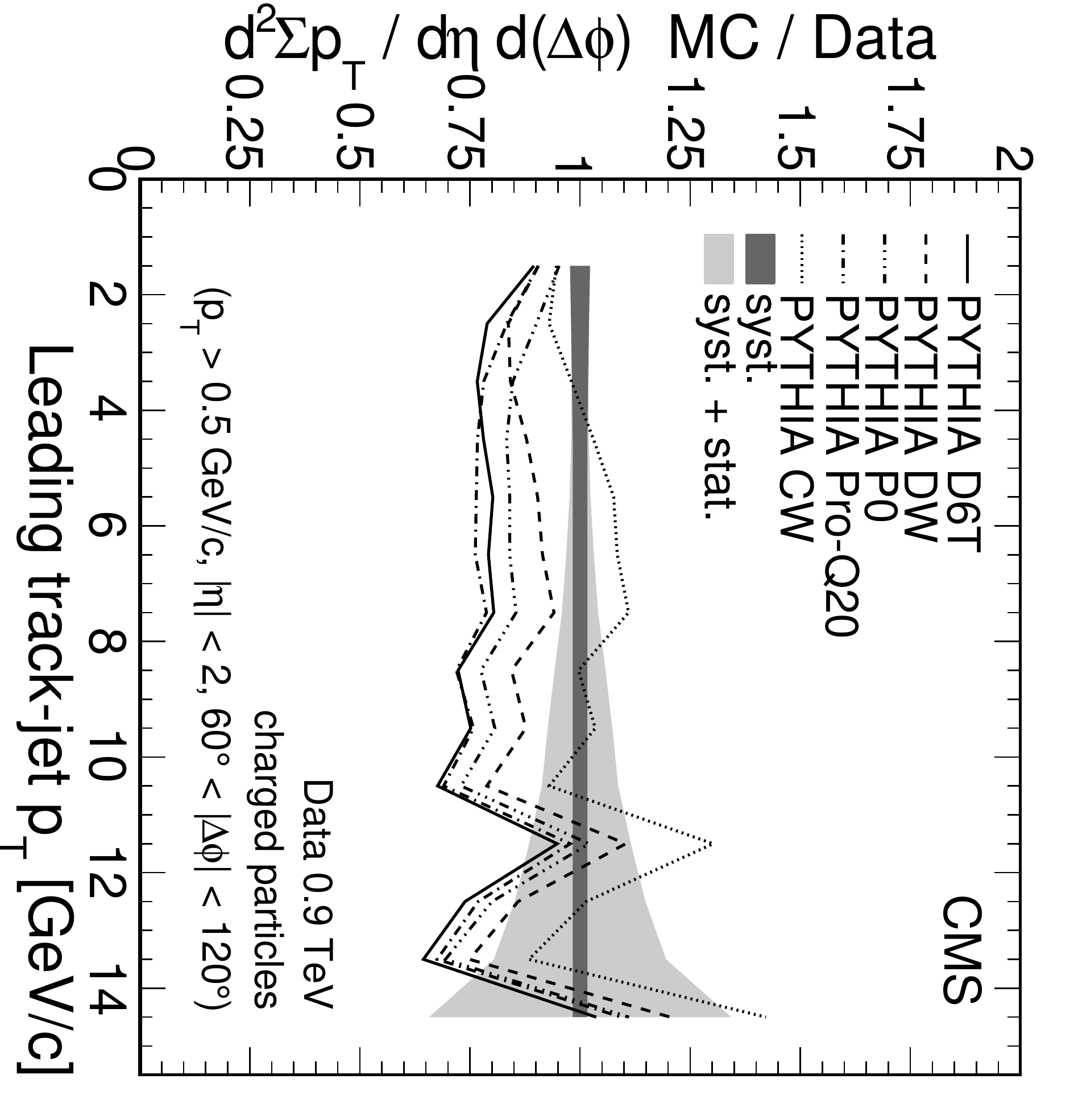} \\
\includegraphics[angle=90, width=0.45 \textwidth]{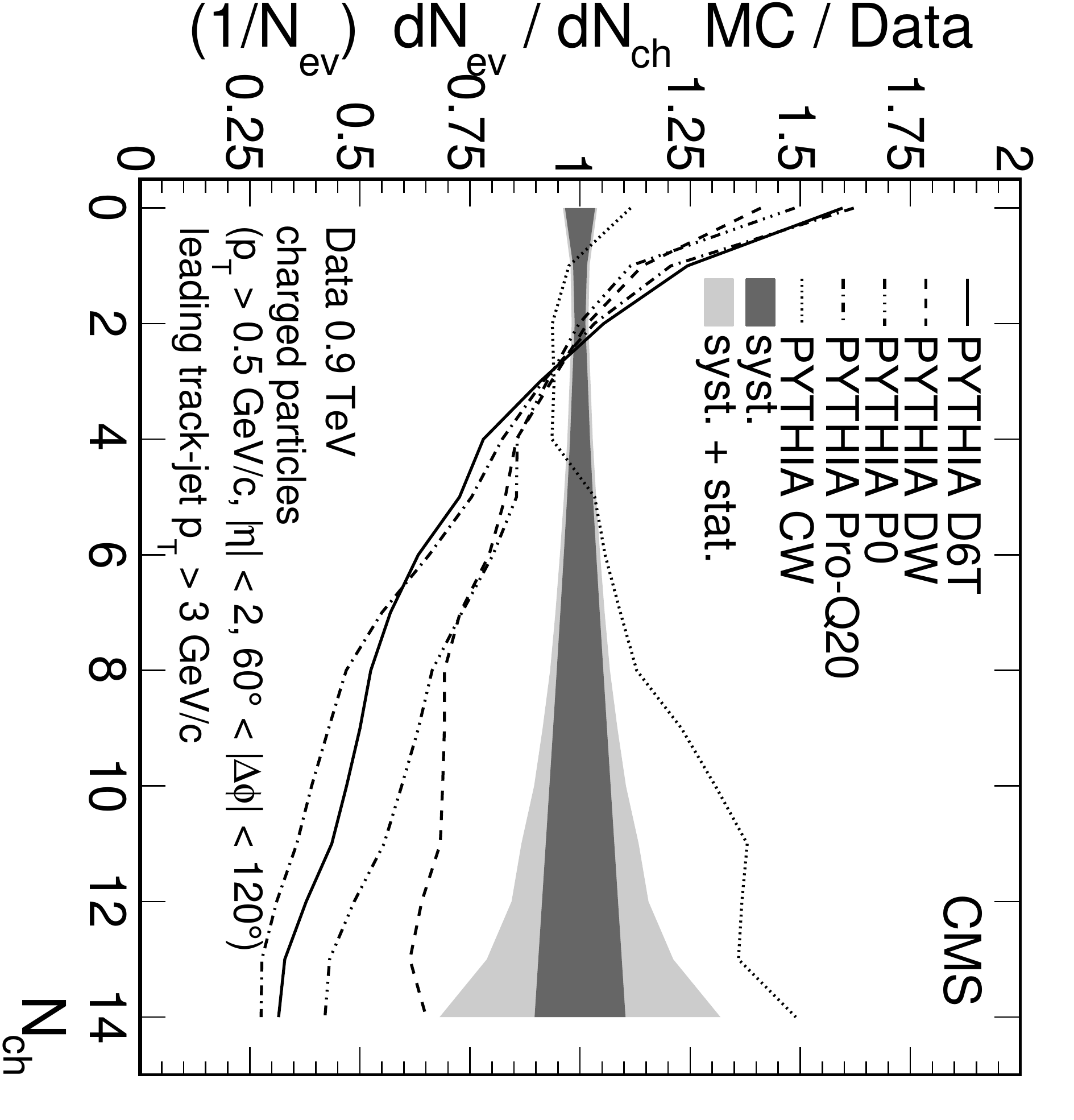}
\includegraphics[angle=90, width=0.45 \textwidth]{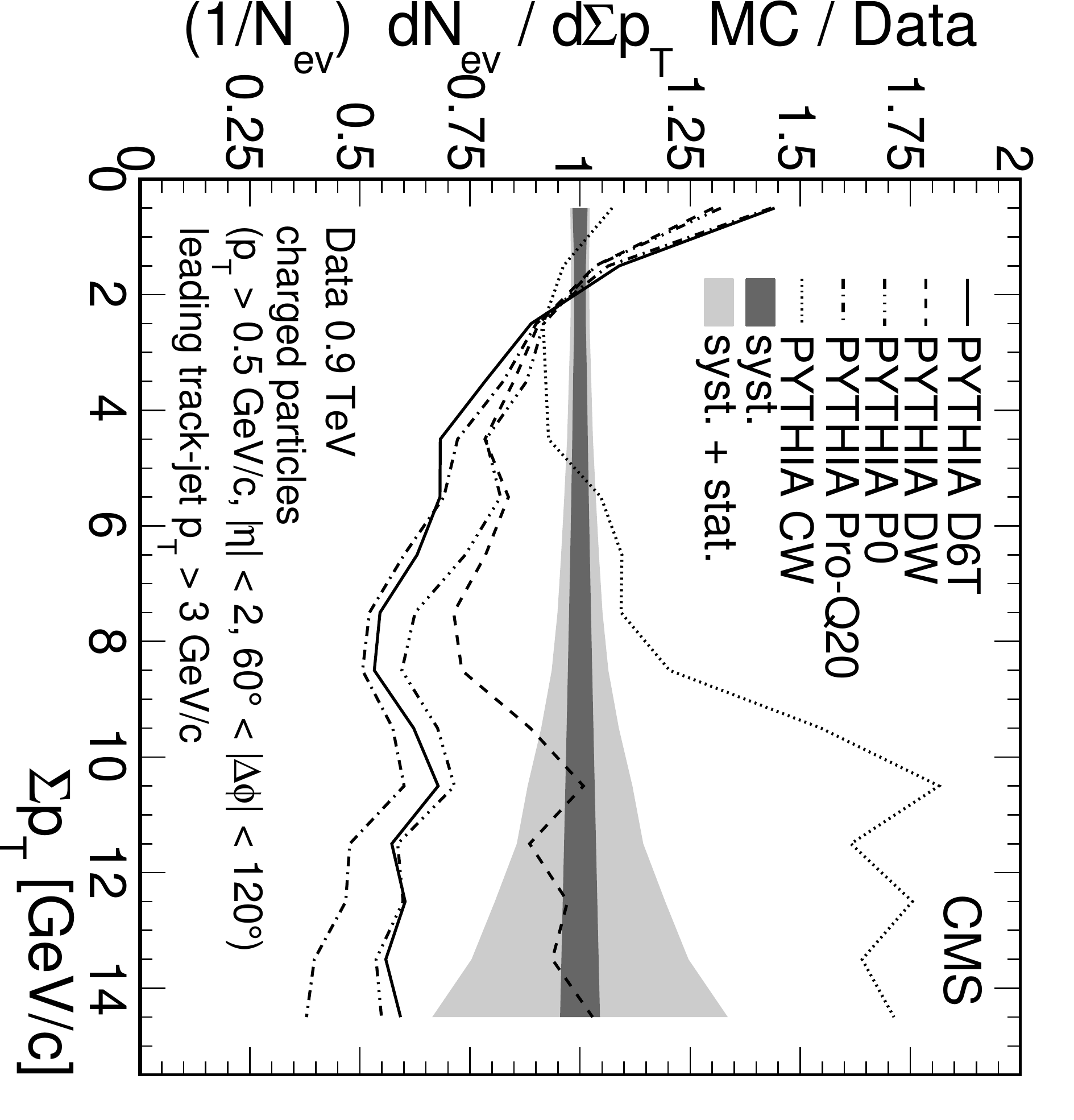}\\
\includegraphics[angle=90, width=0.45 \textwidth]{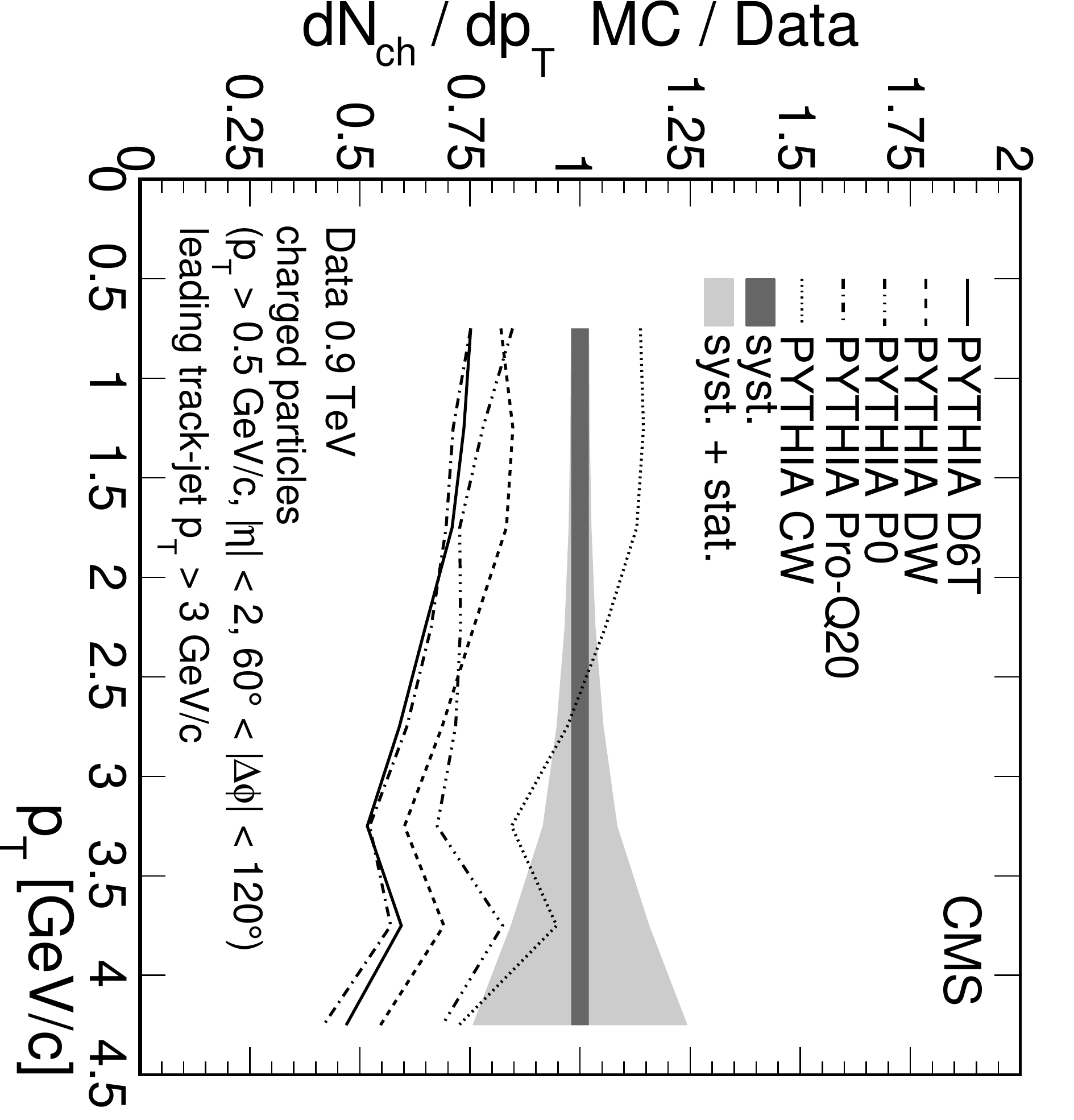}
\end{center}
\caption{
Ratios of various MC predictions, including full detector simulation, to the measurements 
of hadrons with \ptcut\ and \etacut\ in the transverse region, $60^\circ\!<\! |\Delta\phi| \!<\! 120^\circ$:
(from top left to bottom)
average multiplicity of charged particles, as a function of the leading track-jet $p_T$ 
(cf.~Fig.~\ref{fig:CWDWToData}, upper right); 
average scalar $\sum p_T$, as a function of the leading track-jet $p_T$
(cf.~Fig.~\ref{fig:CWDWToData}, lower right); 
distribution of the charged particle multiplicity 
(cf.~Fig.~\ref{fig:diststoDWCW}, upper left); 
distribution of the scalar $\sum p_T$
(cf.~Fig.~\ref{fig:diststoDWCW}, upper right); 
$p_T$ spectrum
(cf.~Fig.~\ref{fig:diststoDWCW}, bottom). 
The inner bands correspond to the systematic uncertainties and the outer bands 
to the total experimental uncertainty (systematic and statistical uncertainties
added in quadrature).
\label{fig:ratiosToMany} }
\end{figure}

The information is summarized in Fig.~\ref{fig:ratiosToMany}, which 
presents the ratio of the MC predictions to the measurements, for the 
variables presented in Figs.~\ref{fig:CWDWToData} 
and~\ref{fig:diststoDWCW}.
The shape of the steeply falling hadron $p_T$ spectrum 
is well described by all tunes, in particular the P0 tune, which achieves good
agreement in the high-momentum tail because of its hard $ p_T$ spectrum.
The CW and DW tunes globally
describe the measurement of hadron production in the transverse region best, 
both in normalization and in shape, with the CW predictions generally higher 
than the data and the DW predictions lower.
A small dependence on the choice of the leading object is observed,
with a preference for CW in the case of a leading track-jet and for DW in the 
case of a leading particle (not shown).
The predictions of tune D6T are too low and generally the least 
consistent with the data. 
The predictions of tunes Pro-Q20 and P0 tend to lie between 
the predictions of tunes D6T and DW.

\section{Summary and Conclusions}
\label{sec:Conclusions}

This paper describes a study of the production of hadrons
with \ptcut\ and \etacut\ at the LHC, in proton-proton collisions
at $\sqrt{s} = 0.9 \TeV$.
Event selection required the presence of a hard scale,
provided by the transverse momentum of the leading charged 
particle or of the leading track-jet.
The minimal value of the scale was chosen  in the range $1$ to $3\GeVc$.
Particular attention has been devoted to the transverse region, defined by the 
difference in azimuthal angle between the leading object and charged particle 
directions,  $60^\circ\!<\! |\Delta\phi| \!<\! 120^\circ$, which is most appropriate 
for the study of the underlying event.

The predictions of several PYTHIA MC models, after full detector simulation, have been 
compared to the data.
The models are all consistent with data taken at the Tevatron at $\sqrt{s} =1.8 \TeV$,
 but they differ 
in the implementation of radiation, fragmentation, and multiple parton interactions.
They describe general features of the data.
In the transverse region most tunes predict too little hadronic activity.
An important parameter of simulation tuning in the PYTHIA framework is the
centre-of-mass energy dependence of the low $\hat{p}_{T_0}$ cutoff aimed at regularizing
singularities in hard scattering and MPI.
The present data favour an energy dependence of this parameter 
along the lines of PYTHIA tune DW ($\epsilon = 0.25$) or even stronger
($\epsilon = 0.30$, as in tune CW).
Lower values of $\epsilon$, as in tune D6T ($\epsilon = 0.16$), are disfavoured. 

The present measurements, together with results from $ {\rm Sp \bar p S}$, Tevatron,
and RHIC, as well as future LHC results at $\sqrt{s} =7$ and $14 \TeV$, are expected to help 
in understanding better the properties of the underlying event and of multiple 
parton interactions in hadron-hadron scattering at high energy.
This is essential for precision measurements of Standard Model processes and for
the search for new physics at the LHC.

\section*{Acknowledgments}

We wish to congratulate our colleagues in the CERN accelerator departments for the excellent performance of the LHC machine. 
We thank the technical and administrative staff at CERN and other CMS institutes for their devoted efforts during the design, construction and operation of CMS. 
The cost of the detectors, computing infrastructure, data acquisition and all other systems without which CMS would not be able to operate was supported by the financing agencies involved in the experiment. 
We are particularly indebted to: the Austrian Federal Ministry 
of Science and Research; the Belgium Fonds de la Recherche Scientifique, and Fonds voor Wetenschappelijk Onderzoek;
the Brazilian Funding Agencies (CNPq, CAPES, FAPERJ, and FAPESP); the Bulgarian Ministry of Education and Science; CERN; 
the Chinese Academy of Sciences, Ministry of Science and Technology, and National Natural Science Foundation of China; 
the Colombian Funding Agency (COLCIENCIAS); the Croatian Ministry of Science, Education and Sport; 
the Research Promotion Foundation, Cyprus; the Estonian Academy of Sciences and NICPB; 
the Academy of Finland, Finnish Ministry of Education, and Helsinki Institute of Physics; 
the Institut National de Physique Nucl\'eaire et de Physique des Particules~/~CNRS, and Commissariat \`a l'\'Energie Atomique, France; 
the Bundesministerium f\"ur Bildung und Forschung, Deutsche Forschungsgemeinschaft, and Helmholtz-Gemeinschaft Deutscher For-schungszentren, Germany; 
the General Secretariat for Research and Technology, Greece; the National Scientific Research Foundation, and National Office for Research and Technology, Hungary;
 the Department of Atomic Energy, and Department of Science and Technology, India; the Institute for Studies in Theoretical Physics and Mathematics, Iran; the Science Foundation, Ireland; 
the Istituto Nazionale di Fisica Nucleare, Italy; the Korean Ministry of Education, Science and Technology and the World Class University program of NRF, Korea; 
the Lithuanian Academy of Sciences; the Mexican Funding Agencies (CINVESTAV, CONACYT, SEP, and UASLP-FAI); the Pakistan Atomic Energy Commission; 
the State Commission for Scientific Research, Poland; the Funda\c{c}\~ao para a Ci\^encia e a Tecnologia, Portugal; JINR (Armenia, Belarus, Georgia, Ukraine, Uzbekistan); 
the Ministry of Science and Technologies of the Russian Federation, and Russian Ministry of Atomic Energy; 
the Ministry of Science and Technological Development of Serbia; the Ministerio de Ciencia e Innovacion, and Programa Consolider-Ingenio 2010, Spain; 
the Swiss Funding Agencies (ETH Board, ETH Zurich, PSI, SNF, UniZH, Canton Zurich, and SER); 
the National Science Council, Taipei; the Scientific and Technical Research Council of Turkey, and Turkish Atomic Energy Authority; 
the Science and Technology Facilities Council, UK; the US Department of Energy, and the US National Science Foundation.
Individuals have received support from the Marie-Curie IEF program (European Union); the Leventis Foundation; the A. P. Sloan Foundation; 
the Alexander von Humboldt Foundation; and the Associazione per lo Sviluppo Scientifico e Tecnologico del Piemonte (Italy).

\clearpage

\bibliography{auto_generated}   

\cleardoublepage\appendix\section{The CMS Collaboration \label{app:collab}}\begin{sloppypar}\hyphenpenalty=5000\widowpenalty=500\clubpenalty=5000\textbf{Yerevan Physics Institute,  Yerevan,  Armenia}\\*[0pt]
V.~Khachatryan, A.M.~Sirunyan, A.~Tumasyan
\vskip\cmsinstskip
\textbf{Institut f\"{u}r Hochenergiephysik der OeAW,  Wien,  Austria}\\*[0pt]
W.~Adam, T.~Bergauer, M.~Dragicevic, J.~Er\"{o}, C.~Fabjan, M.~Friedl, R.~Fr\"{u}hwirth, V.M.~Ghete, J.~Hammer\cmsAuthorMark{1}, S.~H\"{a}nsel, M.~Hoch, N.~H\"{o}rmann, J.~Hrubec, M.~Jeitler, G.~Kasieczka, W.~Kiesenhofer, M.~Krammer, D.~Liko, I.~Mikulec, M.~Pernicka, H.~Rohringer, R.~Sch\"{o}fbeck, J.~Strauss, A.~Taurok, F.~Teischinger, W.~Waltenberger, G.~Walzel, E.~Widl, C.-E.~Wulz
% ...
\vskip\cmsinstskip
% ...
1:~~Also at CERN, European Organization for Nuclear Research, Geneva, Switzerland\\
% ...

\end{sloppypar}
\end{document}